\documentclass{emulateapj}

\def\cmtres{\mbox{cm$^{-3}$}}
\def\cv{\mbox{$c_{\rm vir}$}}
\def\epsSN{\mbox{$\epsilon_{>8}$}}  
\def\etaE{\mbox{$\eta_{\rm >8}$}}  
\def\esf{\mbox{{$\epsilon_{\rm SF}$}}}
\def\erg{\mbox{erg}}

\def\fbU{\mbox{$f_{b,U}$}} 
\def\gammaE{\mbox{$\gamma_{\rm in}$}}  
\def\grav{\mbox{G}}
\def\Ke{\mbox{K}}
\def\kms{\mbox{km/s}}
\def\kpc{\mbox{kpc}}
\def\kpch{\mbox{kpc/h}}
\def\lcdm{{$\Lambda$CDM}}

\def\mg{\mbox{$M_{g,cold}$}}
\def\mgh{\mbox{$M_{g,hot}$}}
\def\mpch{\mbox{$h^{-1}$Mpc}}

\def\ms{\mbox{$M_{s}$}}
\def\msun{\mbox{$M_\odot$}}
\def\msunh{\mbox{$h^{-1}$M$_\odot$}} 
\def\mv{\mbox{$M_{\rm vir}$}}
\def\nsf{\mbox{$n_{\rm SF}$}}

\def\ome{\mbox{$\Omega_m$}}
\def\omel{\mbox{$\Omega_\Lambda$}}
\def\omeb{\mbox{$\Omega_b$}}
\def\rgal{\mbox{$R_{\rm gal}$}}
\def\rhosf{\mbox{$\rho_{\rm SF}$}} 

\def\rhalf{\mbox{$R_{1/2}$}}
\def\re{\mbox{$R_{1/2}$}}

\def\rv{\mbox{$R_{\rm vir}$}}
\def\Tsf{\mbox{{$T_{\rm SF}$}}}

\def\zhalf{\mbox{$z_{half}$}}

\def\mathnew{\mathsurround=0pt} 
\def\simov#1#2{\lower .5pt\vbox{\baselineskip0pt 
    \lineskip-.5pt\ialign{$\mathnew#1\hfil##\hfil$\crcr#2\crcr\sim\crcr}}}   
\def\simgreat{\mathrel{\mathpalette\simov >}}  
\def\'#1{\ifx#1i{\accent"13\i}\else{\accent"13#1}\fi}  

\begin{document}

\title{Cosmological simulations of Milky Way-sized galaxies}

\author{Pedro Col\'in$^1$, Vladimir Avila-Reese$^2$, Santi Roca-F\`abrega$^{3}$ and Octavio Valenzuela $^{2}$}
\affil{$^1$Instituto de Radioastronom\'ia y Astrof\'isica, Universidad Nacional Aut\'onoma de M\'exico, A.P. 72-3 
(Xangari), Morelia, Michoac\'an 58089, M\'exico. $^{2}$ Instituto de Astronom\'ia, Universidad Nacional 
Aut\'onoma de M\'exico, A.P. 70-264, 04510, M\'exico, D.F.; Ciudad Universitaria, D.F., M\'exico.
$^3$}

\begin{abstract}

We introduce a new set of eight Milky Way-sized cosmological simulations performed
using the AMR code ART + Hydrodynamics in a $\Lambda$CDM cosmology.
The set of zoom-in simulations covers present-day virial masses that range 
from $8.3 \times 10^{11} \msun$ to $1.56 \times 10^{12} \msun$ and 
is carried out with our {\it simple} but {\it effective} deterministic star 
formation (SF) and ``explosive'' stellar feedback prescriptions. 
The work is focused on showing the goodness of the simulated set of ``field'' Milky Way-sized galaxies.
To this end, we compare some of the predicted physical 
quantities with the corresponding observed ones. 
Our results are as follows. (a) In agreement with some
previous works, we found circular velocity curves
that are flat or slightly peaked. (b) All simulated galaxies with a significant disk component 
are consistent with the observed Tully-Fisher, radius-mass, and cold gas-stellar mass 
correlations of late-type galaxies. (c) The  disk-dominated galaxies have stellar specific angular 
momenta in agreement with those of late-type galaxies, with values around $10^3$ km/s/kpc.  
(d) The SF rates at $z = 0$ of all runs but one are comparable to those estimated
for the star-forming galaxies. 
(e) The two most spheroid-dominated galaxies formed in halos with late
active merger histories and late bursts of SF, but the other run that ends also as 
dominated by an spheroid, never had major mergers. 
(f) The simulated galaxies lie in the semi-empirical stellar-to-halo mass correlation of local central
galaxies, and those that end up as disk dominated, evolve mostly along the low-mass branch of this correlation. 
Moreover, the baryonic and stellar mass growth histories of these galaxies are proportional to their halo 
mass growth histories since the last 6.5--10 Gyr. (g) Within the virial radii of the simulations, $\approx 25-50\%$ of the 
baryons are missed; the amount of gas in the halo is similar to the one in stars in the galaxy, and most
of this gas is in the warm-hot phase.
  (h) The $z\sim 0$ vertical gas velocity dispersion profiles, $\sigma_z$($r$),
are nearly flat and can be mostly explained by the kinetic energy injected by stars. The average values of $\sigma_z$ 
increase at higher redshifts, following roughly the shape of the SF history. 

\end{abstract}


\section{Introduction}

The galaxy formation and evolution, circumscribed to the hierarchical structure 
formation scenario, is a fascinating problem.
It is a complex phenomenon that involves many physical processes and scales, 
from the formation of the dark and gaseous halos at scales of tens to hundreds of kpc
to the formation of stars in giant molecular clouds at scales of
dozens of pc \citep[e.g.,][]{McKee-Ostriker2007}, passing through the formation of supermassive
black holes at the center of massive galaxies at scales of $10^{-3}$ pc
or less  \citep[e.g.,][]{Volonteri2010}, with their corresponding stage of active  
galactic nuclei (AGN). They are accompanied by a variety of large-scale 
feedback effects such as, for example, supernovae (SNe) explosions and AGN outflows. 
Yet, we can encounter larger scales and 
new phenomena related to galaxy formation during the formation of 
groups and clusters of galaxies
\citep[e.g.,][]{Kravtsov-Borgani2012}. 

It has been long known that disks, as observed, should form
when gas cools, condenses and collapses, within dark matter halos, 
while conserving its angular momentum, obtained through external torques 
\citep{WR1978, FE1980, Mo+1998, Firmani+2000}.  Yet,
until recently, forming extended disk galaxies with flat rotation 
curves in hydrodynamic
simulations, in a hierarchical cosmogony, was a challenge 
\citep[e.g.,][]{Mayer+2008}.

Earlier works show that disks could be produced in simulations
without difficulty but they inevitably ended up 
with too little (specific) angular momentum and too much 
stellar mass at the center of the galaxy
\citep{NB1991, KG1991, NW1994, NFW1995, Somer-Larsen+1999}.
The problem was that, because of an inefficient feedback
or poor resolution or both, ``clumps'' (composed of gas,
stars and dark matter) were too much concentrated by the
time they were accreted by a halo/galaxy. The 
result, at the end, was that this lumpy mass lost most of their 
orbital angular momentum by a physical process called 
dynamical friction \citep[e.g.,][]{NB1991}. Later works, 
with better resolution and/or a more effective stellar feedback improved on
this by producing more extended disks and less massive spheroids 
\citep{Abadi+2003, Somer-Larsen+2003, Governato+2004}.

The relatively recent success on forming realistic galaxies
can be attributed to some degree to resolution but
mainly to a better understanding of the processes that 
play a major role in the classical {\it overcooling} 
\citep{WR1978, DS1986, WF1991} and {\it angular momentum} \citep[e.g.,][]{Mayer+2008}
problems. Since the pioneer works of, for example,
\citet{NB1991} and \citet{NW1994}, it was clear that to avoid 
transforming most of the
gas into stars a kind of stellar feedback was needed. This
source of energy (and momentum) not only could avoid the
overcooling of the gas but it could also solve the angular momentum problem
by puffing the gas in the clumps up, making it less susceptible to 
the loss of angular momentum. It was soon realized that in order 
to do its job the feedback needs to be efficient. Most of
works in the recent years get this in one way or another
\citep{Guedes+2011, Brook+2012, Governato+2012, Agertz+2013,
Hopkins+2014}. For example, in a number of works the
thermal feedback becomes efficient by delaying artificially, for about $\sim 10^7$yr,
the cooling of the gas that surrounds the newly 
formed stellar particle \citep[e.g.,][]{TC2000, Stinson+2006, Governato+2007, Agertz+2011,
Colin+2010, PS2011, Guedes+2011, Teyssier+2013}. In this recipe, further star
formation (SF) is stopped by the high temperatures and low densities
attained by the gas, as a result of turning off the cooling.

Another way of making the feedback efficient is by 
depositing momentum (kinetic energy) directly into the gas 
which, unlike the thermal energy, can not be
radiated away \citep[][]{NW1993, SH2003, Scannapieco+2006, OD2008, 
Marinacci+2014}. This wind method is somehow artificial in the
sense that the kick is put by hand and because the wind particles 
are  temporary decoupled from the hydrodynamic interaction. 
Traditionally, stellar feedback has been associated with the
injection of energy only by SNe and in some
cases with both SNe and stellar wind by massive stars 
\citep{Somer-Larsen+2003, Kravtsov03}, but very recently 
some kind of ``early'' feedback has also been incorporated
\citep{Stinson+2013, Hopkins+2014, Trujillo-Gomez+2015}. This
feedback begins few millions of years before the first  SN explodes
and it includes radiation pressure \citep{Krumholz+2014} and 
photoheating by the ionizing radiation of massive stars.
This latter has been shown can significantly affect the
structure of molecular clouds and in some cases destroy
them \citep{Walch+2012, Colin+2013, Lopez+2014}. On the other hand,
results from \citet{Trujillo-Gomez+2015} show
that radiation pressure alone has a small effect on the 
total stellar mass content, but see \citet{Hopkins+2014}.
Nowadays these effects and others, such us cosmic rays \citep{Salem+2014}
and turbulence in molecular clouds, are slowly being
incorporated in the simulations of galaxy formation. 

The MW-sized galaxies are special in the sense that they occupy the peak 
of the stellar mass growth efficiency within dark matter (DM) halos measured
through the $\ms/\mv$ fraction \citep[e.g.,][]{Avila-Reese+2011b}. 
For less massive galaxies, this efficiency decreases because of the global effects 
of SN-driven outflows, and for more massive ones, because of the long cooling times of the shock
heated gas during the virialization of massive halos and the feedback from luminous AGNs
\citep[for a recent review on these processes see][]{Somerville-Dave2014}. 
Therefore, since the baryonic mass assembly of MW-sized galaxies is less 
affected by these astrophysical processes, the shape of their mass assembly history is expected
not to deviate dramatically from the way their halos are assembled \citep[see e.g.,][]{Conroy+2009,Firmani+2010,
Behroozi+2013,Moster+2013}, being then interesting objects to constrain the $\Lambda$CDM cosmology.
Moreover, being these galaxies less susceptible to the large-scale effects of stellar and AGN feedback, 
they are optimal to probe simple models of large-scale SF and self-regulation, where 
the generation of turbulence (traced by the cold gas velocity dispersion) and its
dissipation in the ISM are key ingredients.

The aim of this work is to introduce a new  suite of zoom-in Milky Way (MW) 
sized simulations run with the
 N-body + Hydrodynamic ART code \citep{Kravtsov03} and with the
SF and stellar feedback prescription
implemented for the first time in 
\cite[][see also Avila-Reese et al. 2011a]{Colin+2010}, and used recently in
\citet{Santi+2016} for the ``Garrotxa'' simulations.  
Here we use the same \nsf\ and \esf\ values as in the latter paper. 
The dark matter (DM) halos were drawn from a 50 \mpch\ on a side box and
have each about 1 million DM particles and cells within their virial radii.
The spatial resolution, the side of the cell in the
maximum level of refinement, is 136 pc. The runs use a
deterministic SF approach; that is, stellar particles form {\it every} 
time\footnote{The timestep for star formation is given by the 
timestep of the root grid which varies from about 10 Myr at high $z$ to
$\sim 40$ Myr at $z \sim 0$.}
 gas satisfies certain conditions, with a conversion of
gas to stars of 65\%. Immediately after they are born, they
dump $2 \times 10^{51}$ ergs of thermal energy, for every star 
$> 8\msun$, to the gas in the
cell where the particle is located, rising its temperature to several
$10^7\ \Ke$. This high temperature value comes from 
the high SF efficiency we have assumed and from the fact that all
the injected energy is concentrated in time and space
(see subsection \ref{feedback}).
We sometimes call this form of stellar feedback ``explosive'',
as opposed to the typical assumed feedback consisting of a 
continuous supply of thermal energy for about 40 Myr \citep[e.g.,][]{CK2009}.
At this high temperature, the cooling time is much longer than the 
crossing time \citep{DS2012} and so, the differences in the
galaxies simulated under the assumption of a 
delayed cooling and those that do not consider it, 
are not expected to be significant. 

From our sample of eight simulated halos/galaxies, we distinguish four that
are clearly disk dominated. Two more have kinematic  bulge-to-disk (B/D)
ratios of 1.3 and 2.3. The other two had a violent late
assembly phase and ended up with B/D ratios of 3 and 10.
Here, we study the evolution of the specific angular momentum
and the stellar-to-halo mass fraction, $\ms/\mv$, as a function
of stellar mass. As usual in this kind of studies, we also compute 
the circular velocity profiles and the  SF 
histories. The MW-sized galaxies simulated here present 
realistic $\ms/\mv$ fractions and
are consistent with several observational correlations.  
We find, in agreement with our previous studies
\citep[e.g.,][]{Gonzalez+2014}, nearly flat circular
velocity curves. However, contrary to our results on the
low-mass galaxies \citep{Avila-Reese+2011, Gonzalez+2014},
and to some degree also on the MW-sized galaxy of 
\citet{Santi+2016}, here our
predicted $\ms/\mv$ values agree with those determined by
semi-empirical methods. 

The suite of MW-sized galaxies introduced here will be used elsewhere to study in 
detail the spatially-resolved SF and stellar mass growth histories with the 
aim to compare the results with look-back time studies and fossil record inferences 
from observational surveys like MaNGA/SDSS-IV \citep{Bundy+2015}.
They also will be used to study observational biases after post-processing 
them to include dust and spectral energy distributions to each stellar particle, and 
by further performing mock observations.   

This paper is organized as follows. In Section~\ref{sec:model}, we 
describe the code, the  SF and feedback prescriptions, as well as 
the simulations.  In Section~\ref{sec:results}, we 
put our simulated galaxies into an
observational context by presenting their circular velocity curves,
and by comparing the predicted galaxy properties and correlations with observations.
Then, we estimate the morphology of our simulated galaxies 
and present the spatial distribution of light in some color bands. 
Finally, in the rest of the Section~\ref{sec:results} we present: (i) the 
halo mass aggregation and specific angular momentum evolution of the runs, (ii)
the \ms/\mv\ ratios and their evolution, (iii) the gas-to-stellar
mass ratios, and (iv) the cold gas velocity dispersion profiles. 
In Section~\ref{sec:discussion}, we discuss the resolution issue and
the validity our ``explosive'' stellar feedback. We also discuss
the implications of the the dark/stellar mass assembly 
of our simulated galaxies, as well as the role of the SN 
feedback as the driver of turbulence and self-regulated SF
in the disk.
Our conclusions are given in in Section~\ref{conclusions}.

\section{Numerical Methodology} \label{sec:model}

\subsection{The Code} \label{sec:code}

The numerical simulations used in this work were run using 
the N-body + hydrodynamic Adaptive Refinement Tree (ART) code  
\citep[]{KKK97,Kravtsov03}. The code incorporates a variety
of physical processes that are essential to the modeling of 
galaxy formation such as gas cooling, SF, 
stellar feedback, advection of metals, and a UV heating background source.
The Compton heating/cooling, atomic and molecular 
cooling, and UV heating from a cosmological background radiation \citep{HM96}, are all
included in the computation of the cooling/heating rates. These
are tabulated for a temperature range of $10^2 < T < 10^9\ \Ke$ and a grid of densities,
metallicities, and redshifts using the CLOUDY code \citep[version 96b4]{Ferland98}.

\subsection{Star formation and stellar feedback}

Since the subgrid physics of SF and stellar feedback implemented in the
code is discussed in detail in \citet{Colin+2010} 
\citep[see also][]{Avila-Reese+2011,Colin+2013,Colin+2015,Santi+2016},
here we only give a brief summary. In the code, a cell form
a stellar particle (hereafter SP, unless otherwise stated) 
if they are dense and cool enough; that is, if
$T < \Tsf$ and $\rho_g > \rhosf$, where $T$ and $\rho_g$ are the temperature and 
density of the gas, respectively, and $\Tsf$ and $\rhosf$ are the temperature and
density threshold, respectively. Here, we use the same values of 
$\Tsf$ and $\nsf$ parameters as in \citet{Santi+2016}; namely, 9000 \Ke\ and 1 \cmtres,
respectively, where \nsf\ is the density threshold in hydrogen atoms per
cubic centimeter. In practice, the temperature requirement is almost always 
satisfied once the gas density reaches a value above $\rhosf$. In
the present scheme of our subgrid SF+feedback recipe the outcome of the simulation
is sensitive to $\nsf$, the higher it is the less efficient is the stellar
feedback; see \citet{Colin+2010} for a discussion on the choice of the
$\nsf$ values. 
A stellar particle of mass $m_* = \esf\ m_g$ is placed in a 
grid cell every time the above conditions are simultaneously satisfied, 
where $m_g$ is the gas mass in the cell and \esf\ is a parameter that measures 
the local efficiency by which gas is converted into stars. As in \citet{Santi+2016}, 
we set $\esf = 0.65$. 
The reason why we choose this value, 0.65, is because it was found in \citet{Santi+2016}
to be optimal within the context of our SF-feedback recipe, given the spatial resolution
attained in our simulations. Our aim here is to investigate how fair the simulations
with the mentioned subgrid parameters are in terms of producing realistic results.

For the stellar feedback, we use what we called the ``explosive'' stellar thermal feedback recipe, 
according to which each SP injects, just immediately after formation, into the cell 
$E_{{\rm SN+Wind}} = 2 \times 10^{51}\ \erg$ of thermal energy {\it for each star more massive 
than  8 \msun}. Half of this energy is assumed to come from Type II SN 
and half from shocked stellar winds. This thermal energy dumped into just one cell
and suddenly is capable of raising the temperature of the cell to values 
$\simgreat 10^7\ \Ke$; the precise value depends on the assumed initial mass function (IMF),
the assumed $E_{{\rm SN+Wind}}$ value, and the value of the \esf\ parameter
(see also Section \ref{SFandFeedback}). 
On the other hand, each $8 \msun$ ejects  $1.3 \msun$ of metals. For the assumed \citet{MS79} IMF, 
a stellar particle of $10^5\ \msun$ produces 749 Type II SNe. Although $2 \times 10^{51}\ \erg$
for the energy of a SN (plus the energy provided by the stellar wind) is certainly
an upper limit, we note that a change of the \citet{MS79} IMF for 
a Chabrier-like IMF \citep[][]{Chabrier2005} will produce a 
factor of $\simgreat 2$ more massive stars, with the corresponding augment in thermal energy.

Although in this set of simulations we still turn off the cooling for some time, 
40 Myr, in those cells where young stellar particles are located, tests done
with low-mass halos/galaxies show that delaying the cooling for
few tens of Myr is not expected to
significantly influence our results (see also subsection \ref{feedback}). 
This is because for the typical densities and
temperatures found in the star-forming cells, immediately after the formation
of a stellar particle ($\simgreat 1\ \cmtres$
and $T \simgreat\ 10^7 \Ke$), the cooling time is actually much  longer
than the crossing time \citep{DS2012}; in other words, most of the
times the gas in the cell, where the newborn stellar particle is, will 
expand before radiating away its heat. 

\subsection{The Simulations}

The aim of the paper is to introduce a new set of 
``field'' MW-sized galaxies, simulated within the hierarchical CDM structure
formation scenario, with the hydrodynamic version of the ART code
and the ``explosive'' stellar feedback, and study their structure and evolution.
The simulations are performed
in the \lcdm\ cosmology with $\ome = 0.3$, $\omel = 0.7$, and $\omeb = 0.045$. 
The CDM power spectrum is taken from \citet{kh97} and it is normalized to 
$\sigma_8 = 0.8$, where $\sigma_8$ is the rms amplitude of mass fluctuations in 
8 \mpch\ spheres. These values are close to those inferred from the Planck
collaboration \citep{Planck}.

We choose our eight halos, of present-day masses around $10^{12}\ \msun$, 
from a DM N-body-only ART simulation, run in a box
of 50 Mpc/h on a side with $128^3$ DM particles. 
Outside one halo (run Sp3, see below), 
which has a companion of comparable mass at a distance of 0.26 \mpch, 
all the others
are relatively isolated: they do not have a mate with a mass $\simgreat 
5 \times 10^{11} \msun$ at a distance lower than 1 \mpch.  
Thus, the environment of our galaxies should not be associated to one of groups/clusters and 
can be related to what observers call the field.  
The region, selected to be resimulated with much higher resolution and with the 
N-body+hydrodynamics ART, is always a sphere with a radius of three times
the virial radius\footnote{The virial radius is defined as the radius that encloses a mean 
density equal to $\Delta_{\rm vir}$ times the mean density of the universe, where $\Delta_{\rm vir}$ is 
obtained from the spherical top-hat collapse model. 
For our cosmology, $\Delta_{\rm vir}$=338 at $z=0$.}
 \rv, centered on the selected halo.
The use of a rather large resimulated region keeps the contamination of the 
target halos by massive particles to very low levels. 
The multiple-mass species initial conditions were run
using the code PMstartM \citep{KKBP01}. The Lagrangian region, corresponding to 
a sphere of 3\rv\ radius at $z = 0$, is identified at $z= 100$ 
and resampled with additional small-scale waves
\citep{KKBP01}. The number of DM particles in the high-resolution zone goes
from about 1.5 to 2 million and the mass
per particle $m_p$ is $1.02 \times 10^6\ \msunh$. This resolution is 
comparable to some of the recently published works \cite[e.g.,][]{Marinacci+2014}.

As in previous papers \citep[e.g.,][]{Avila-Reese+2011}, the 
whole computational box is initially covered by a 
uniform grid of $128^3$ zeroth-level cells. The Lagrangian region,
on the other hand, is refined unconditionally to the 
fourth level, corresponding to an effective mesh size of $2048^3$, immediately after
the onset of the simulation. Beyond this level, the grid is 
refined recursively, as the matter
distribution evolves, using as a criteria the DM or gas density. 
In this set of simulations, cells are refined 
when its mass in DM exceeds $5.1m_p$ or the gas mass is greater 
than 5.1 times $\fbU m_p$, where $\fbU = \omeb/\ome$
is the universal baryon fraction. We set the maximum level of refinement to
12 so that the high density regions, where disks are located, are 
mostly filled, at present-day, with cells of 136 pc per side. 
On the other hand, the virial sphere centered
on the galaxy is covered by around 1 million number of resolution elements.

DM halos are identified by a variant of the bound density 
maxima (BDM) halo finder algorithm described in \citet{kh97}. The
code was kindly provided by 
A. Kravtsov. It is run on the DM particles of species 1 (those with
the smallest mass). The galaxies under study are centered at the position of the 
corresponding most massive halos.

\section{Results} \label{sec:results}

\subsection{Galaxy properties at $z = 0$}

As mentioned above, we focus our study on distinct halos with a mass similar to that of
the MW, $\mv\approx 10^{12}\ \msun$ at $z=0$.
The list of runs and some global parameters such as the total mass, \mv,  are presented
in Table~\ref{tab:sims}. Column (1) gives the identification of the simulation, where
Sp stands for ``spiral'' (though, not all runs produce disk dominated
galaxies).
In column (2) we show the virial mass, 
which includes DM and baryonic mass inside \rv, while in column (7) the
peak value of the circular velocity, $V_{\rm max}$, is given. In columns (3) and (4) we show 
the stellar and cold gas mass inside the galaxy radius, \rgal, defined as 
$\rgal = 0.1\rv$, respectively. Cold gas is all gas below 
$1.5 \times 10^4$ K. Although the code has molecular cooling
and the minimum gas temperature is set to 300 K, in practice this low
temperature is almost never attained because there is not self-shielding.
The background UV radiation maintains the cold gas in the disk at several 
thousands degrees. In the next column, the mass of the hot gas, as defined as 
all gas above $3.0 \times 10^5$ K, is shown. The half-mass radius,
\rhalf, of the corresponding run, computed as the radius that encloses half 
of the stellar mass of the galaxy, is shown in column (6). Column (8) shows the
cold gas fraction, $f_g\equiv \mg/(\mg + \ms)$, while in the next two columns, two 
estimates of the kinematic disk-to-total ratio (see below) are presented. 

\begin{deluxetable*}{ccccccccccccc}
\tablecolumns{11}
\tablewidth{0pc}
\tablecaption{Global properties of galaxies at $z=0$}
\tablehead{\colhead{Run} & \colhead{\mv} & 
\colhead{\ms\tablenotemark{a}} & \colhead{\mg\tablenotemark{b}}
& \colhead{\mgh\tablenotemark{c}} & \colhead{\re\tablenotemark{d}} & \colhead{$V_{\rm max}$} & 
\colhead{$f_g$\tablenotemark{e}} & \colhead{D/T\tablenotemark{f}}
& \colhead{$f_{disk}$\tablenotemark{g}} & \colhead{SFR} & $c_{vir}$\tablenotemark{h} 
&  $\delta_{1200}$\tablenotemark{i} \\
 & ($10^{12}$ \msun) & ($10^{10}$ \msun) & ($10^9$ \msun) & ($10^{10}$ \msun) & 
(kpc) & (\kms) &  &  &  & ($\msun/$yr) &  & }
\startdata
Sp1 & 0.84 & 1.8 &  5.4 & 1.9 & 6.0 & 164.5 & 0.23 & 0.75 & 0.65 & 0.9 & 13.2 &  5.3 \\
Sp2 & 0.83 & 3.3 & 11.4 & 1.4 & 4.9 & 195.9 & 0.26 & 0.43 & 0.40 & 0.2 & 22.8 &  4.7 \\
Sp3 & 0.99 & 5.3 &  3.9 & 3.4 & 6.4 & 206.7 & 0.06 & 0.81 & 0.70 & 1.0 & 14.8 &  8.7 \\
Sp4 & 1.56 & 8.4 & 18.7 & 9.9 & 6.8 & 225.4 & 0.18 & 0.25 & 0.26 & 5.6 & 12.6 &  9.9 \\
Sp5 & 1.05 & 4.1 & 4.2  & 5.0 & 3.3 & 207.5 & 0.09 & 0.09 & 0.19 & 1.4 & 13.4 & 11.3 \\
Sp6 & 0.97 & 2.3 & 4.2  & 2.7 & 2.6 & 197.2 & 0.16 & 0.30 & 0.39 & 0.7 & 14.7 &  5.6 \\
Sp7 & 1.09 & 5.4 & 6.4  & 3.3 & 5.2 & 226.1 & 0.11 & 0.69 & 0.62 & 3.2 & 24.7 &  6.2 \\
Sp8 & 1.20 & 6.3 & 11.5 & 4.5 & 5.9 & 219.7 & 0.19 & 0.64 & 0.65 & 4.3 & 15.4 &  7.3 \\
\enddata
\tablenotetext{a}{Stellar mass within $\rgal = 0.1\rv$.} 
\tablenotetext{b}{Mass of cold gas ($T < 1.5 \times 10^4\ \Ke$) inside \rgal.}
\tablenotetext{c}{Mass of hot gas ($T > 3 \times 10^5\ \Ke$) inside \rv.}
\tablenotetext{d}{Radius that encloses half of the stellar mass within \rgal.}
\tablenotetext{e}{$f_g \equiv M_g/(M_g + M_s)$}
\tablenotetext{f}{Ratio of the mass contained in the high-angular momentum disk stars 
with respect togthe stellar mass inside \rgal.}
\tablenotetext{g}{Ratio between the mass of all stellar particles with 
stellar circularities greater than 0.5 and the stellar mass in the galaxy.}
\tablenotetext{h}{$c_{vir}$ is defined as the ratio between \rv\ and $r_s$, where $r_s$ is
computed fitting the DM halo density profile to the NFW parametric function.}
\tablenotetext{i}{The mean overdensity inside a sphere of 1.2 Mpc radius centered 
on the halo.}
\label{tab:sims}
\end{deluxetable*} 

To have more information about the nature of the halos, where 
our set of galaxies are located, besides their present-day masses
we also compute how concentrated they are and
what kind of environment they inhabit. As a measure of concentration, \cv, 
we use the ratio between \rv\ and the scale radius $r_s$. This latter is computed 
by fitting the DM halo density profile of the hydrodynamic simulations to the NFW 
parametric function. On the other hand, as a measure of environment, we
use $\delta_{1200}$ \citep{Creasey+2015}; $\delta_{1200}$ is the mean overdensity inside
the sphere of 1.2 Mpc radius, centered on the halo, computed by
using the low-resolution DM-only simulation. We find \cv\ values that
span from a minimum value of 12.6 obtained by Sp4 to a maximum 
value of 24.7 reached by Sp7. 
As a reference, \citet{Rodriguez-Puebla+2016} report mean values of $\cv\approx 11$ for 
$\mv=10^{12} \msun$ from fits to several large-volume N-body cosmological simulations.
Because of the effects of baryons, the NFW halo concentration is expected to significantly increase
as reported in, e.g., \citet{Santi+2016}. In fact, the shape of the halo density profiles 
(not shown in this paper) changes with
respect to the NFW function, showing a hump below 5-8 kpc  \citep[see also][]{Guedes+2011,Schaller+2016}.
Moreover, halos have $\delta_{1200}$ values that go from 4.7 for Sp2 to 11.3 for Sp5.
Overall, our halo sample is diverse in the sense that they have 
a variety of concentrations and inhabit different overdensity environments.

\begin{figure}[htb!]
\plotone{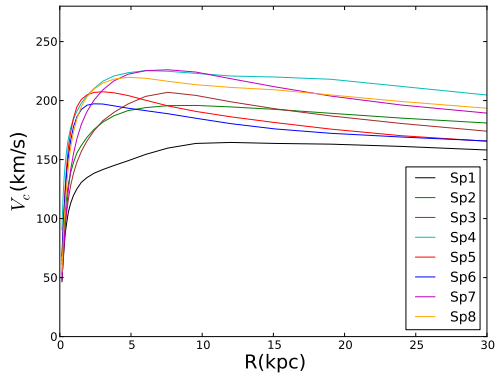} 
\caption[fig:vc]{Circular velocity curves measured as $\sqrt{\grav M(<r)/r}$ for 
our set of runs.  As already found in \citet{Santi+2016} for their high-resolution
simulated galaxy, we found also here curves that do not show a central peak.}
\label{fig:vc}
\end{figure}


\begin{figure}[htb!]
\plotone{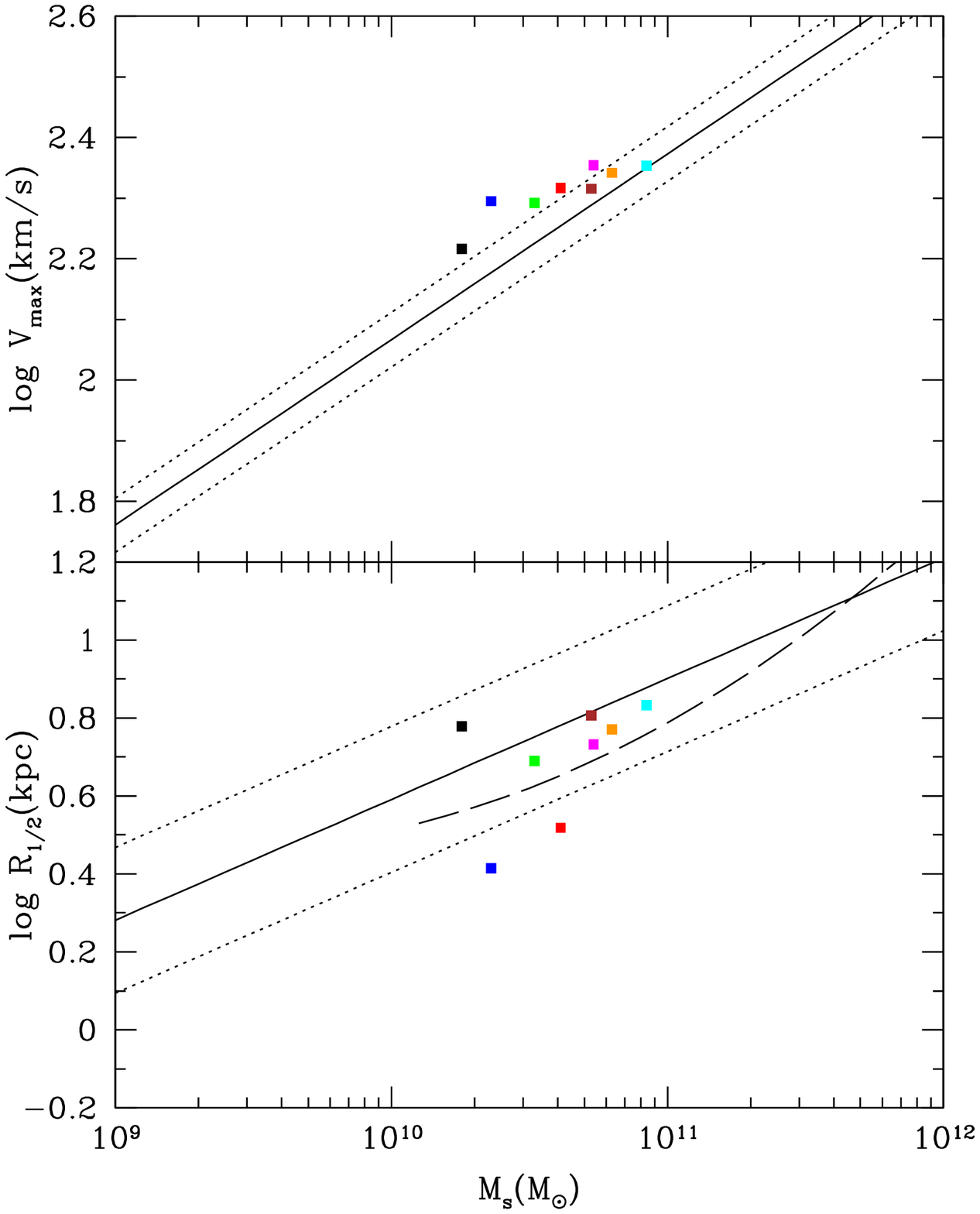} 
\caption[fig:TF]{Maximum circular velocity (top panel) and
half-mass radius (lower panel) are plotted as a function of the galaxy stellar mass 
for our set of simulated galaxies (colored, solid squares). 
The color-code setting is the same as in Figure~\ref{fig:vc}.
The mean $V_{\rm max}-\ms$ (Tully-Fisher) and \re--\ms\ correlations
inferred from observations by \citet{Avila-Reese+2008} 
are shown as solid lines. The dotted lines encompass the estimated 1$\sigma$ intrinsic scatter.
The dashed line in the lower panel is a more recent determination of the  \re--\ms\ relation
of late-type galaxies for a large SDSS sample by \citet{Bernardi+2014}.
\label{fig:TF}}
\end{figure}

\begin{figure*}[htb!]
\plotone{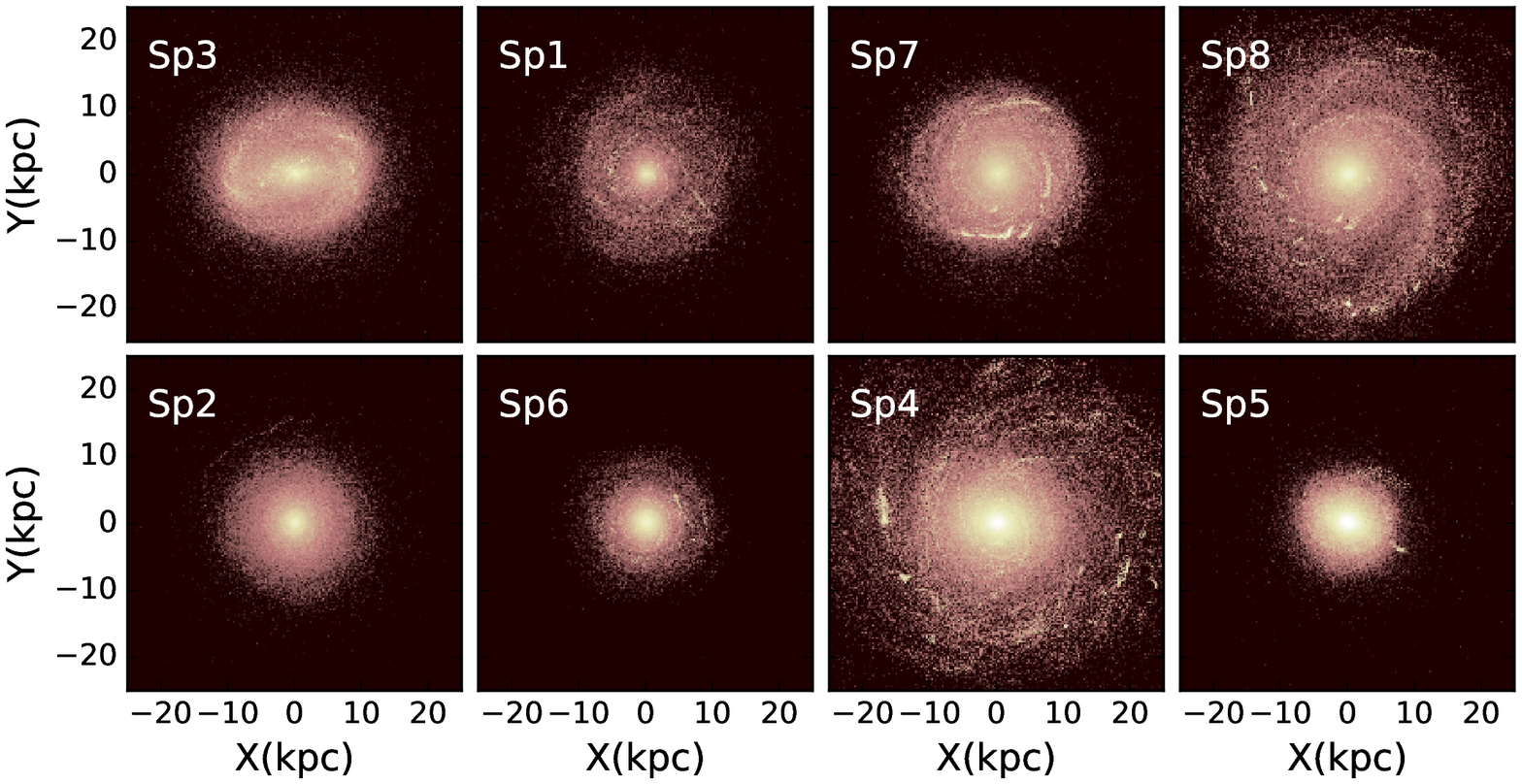} 
\plotone{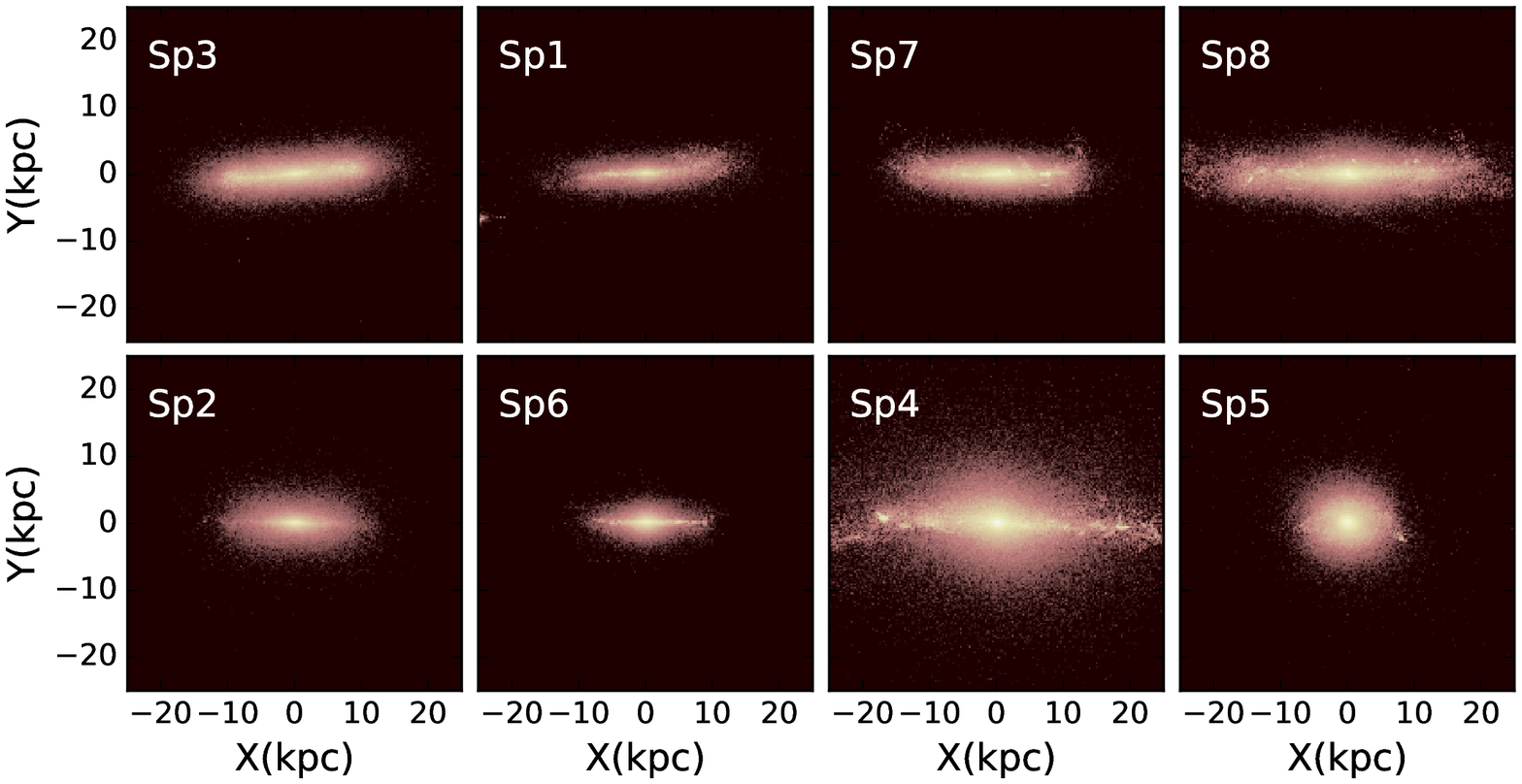} 
\caption[fig:RSD6plots]{Luminosity surface density in the R band in $L_\odot/\kpc^2$ units 
for the eight runs sorted from highest to lowest values of $D/T$. The scale of colors goes from 
5 to 7.2 in the log, the lighter the color the higher the luminosity. {\it Top panels} show a face-on 
view of the galaxies while {\it bottom panels} show the edge-on view. }
\label{fig:RSD8plots}
\end{figure*}

In Figure~\ref{fig:vc}, we depict the circular velocity curves
$\sqrt{G M(<r)/r}$, as a proxy for $V_{\rm rot}$, 
for our entire set of runs. The color setting is shown inside the panel
and will remain so for other figures unless otherwise stated.
In agreement with the results of \citet{Santi+2016}, our curves are also nearly flat. We see, 
however, some differences among the runs. For instance, Sp1 shows a gently rising inner 
part and ends up with the lowest $V_{\rm max}$ 
value; moreover, Sp7 has a $V_c$ curve similar to
that of the MW. At the solar radius, $V_c$ is 224 \kms\ close to
the value accepted for the rotation velocity of the Galaxy \citep[e.g.,][]{Koposov+2010}. 
It is interesting to note that
although the $V_c$ curves of runs Sp5 and Sp6 are very similar (they are slightly peaked), 
the former does not show a stellar disk (see Figure~\ref{fig:RSD8plots} below) whereas Sp6 does.
One then needs to be cautious when trying to infer the presence of a disk by just looking
at the shape of the circular velocity curve.

In the upper panel of Figure~\ref{fig:TF}, we plot the $V_{\rm max}$ values, shown in Table~\ref{tab:sims},
against \ms\ (colored, solid squares). 
The color-code of the runs is the same as in Figure~\ref{fig:vc}. 
The solid black line is the best orthogonal fit to a compilation and homogenization of observations by \citet{Avila-Reese+2008};
the masses were corrected by -0.1 dex to pass from a diet-Salpeter IMF to a Chabrier one. The dotted
lines show the intrinsic scatter ($1\sigma$) reported in that work. 
In general, the simulated galaxies follow the observed stellar Tully-Fisher relation.
In a more detailed comparison, we see that 
three of the disk-dominated runs lie above the $1\sigma$ intrinsic scatter, with a shift of 
$\approx 0.07$ dex in $ \log V_{\rm max}$ from the mean relation.
In the case of the spheroid-dominated galaxies, they are expected to have higher 
values of  $V_{\rm max}$ than the disk-dominated ones. Other authors have also 
found, for their simulated MW-sized galaxies,
a slight shift in the Tully-Fisher relation towards the high velocity side \citep[c.f.][]{Guedes+2011, Aumer+2013,
Marinacci+2014,Murante+2015}. However, biases in the selection process of simulated systems with respect
to the observed galaxy samples, as well as differences between the observational measures and those performed
here, unable us to claim whether there is a potential problem or not.

In the lower panel of Figure~\ref{fig:TF},  \re\ against \ms\ is shown for the eight runs (colored, solid squares).
The solid and dotted lines are the linear fit (in logarithmic scales) to the same galaxy sample used in \citet{Avila-Reese+2008} for 
determining the Tully-Fisher relation. These authors present the galaxy scale radius, so we multiply by a factor of 1.68 to pass
to \re. We also show in Figure~\ref{fig:TF} the mean \re--\ms\ relation as inferred in \citet{Bernardi+2014} for a
large sample of late-type SDSS galaxies and after a careful two-component photometric decomposition of them;
their analysis is only for galaxies more massive than $10^{10}$ \msun.
In general, we see that our simulated galaxies are in good agreement with the local \re--\ms\ correlation from observations.
The two runs that deviate most, Sp5 (red) and Sp6 (blue), are spheroid-dominated galaxies.
Observations show that early-type galaxy have indeed smaller effective radii than 
late-type ones \citep[see e.g.,][]{Bernardi+2014}.

\subsection{Morphology}

\begin{figure}[htb!]
\plotone{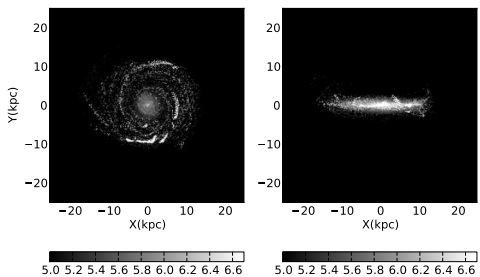} 
\caption[fig:SDuvSp7]{Luminosity surface densities in the U band in $L_\odot/\kpc^2$ units for run Sp7. 
As the U band samples young stellar populations, 
this figure is representative of the places in the galaxy where
SF is going on. The U band traces better the spiral arms and the thin disk than
the R band (compare this figure with the panels corresponding to Sp7 in
Fig. \ref{fig:RSD8plots}}. 
\label{fig:SDuvSp7}
\end{figure}

\begin{figure*}[htb!]
\plotone{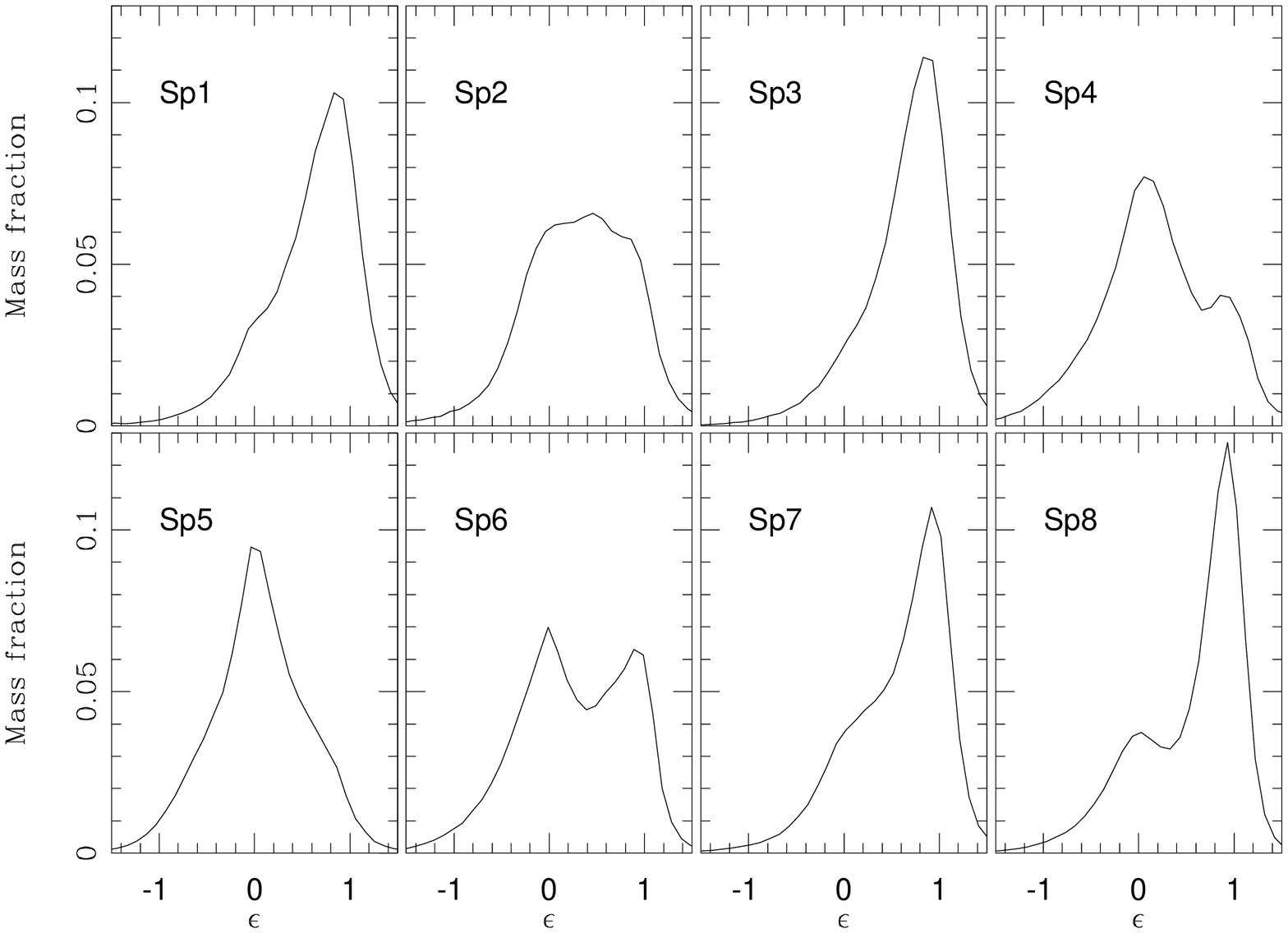} 
\caption[fig:epsall]{Distribution of the mass-weighted stellar circularities
$\epsilon$ for our sample of galaxies at $z = 0$. We have considered only the
stellar particles inside the sphere with radius 0.1\rv\ and have used
the definition by \citet{Scannapieco+2009} in the computation of $j_c$, the specific
angular momentum of a reference circular orbit. As expected from the $D/T$ (or
$f_{disk}$ values) values of Table~\ref{tab:sims} and previous discussion on morphology,
we see that the component associated with the disk of runs Sp4 and Sp5 do not
contribute much to the stellar mass of the galaxy. Sp2 and Sp6, moreover, are 
intermediate cases.}
\label{fig:epsall}
\end{figure*}

In Fig.~\ref{fig:RSD8plots}, we show the spatial luminosity distributions 
in the R band (proportional to the stellar mass distribution)
for our eight simulated galaxies at $z = 0$, both in face-on and edge-on projections. 
The luminosity surface density grid was constructed by
using the publicly available stellar evolution code 
CMD2.7 \citep{Marigo2008}, assuming solar metallicity. All images use the same logarithmic
scale which goes from 5.0 to 7.2 in solar luminosities per kpc$^2$
and cover the same physical extension of 50 kpc of side. At any epoch, in particular 
at $z = 0$, the
plane of the disk, and thus the face-on orientation, is defined 
by the angular momentum of the cold gas inside a sphere of radius
$\sim 5$ \kpch\ comoving. We check that the plane 
orientation is not sensitive 
to the value of this radius\footnote{This is actually achieved by setting the
value for this radius large enough so that the sphere defined by 
it covers a significant fraction of the cold gas of
the galaxy.}. 

Galaxies in Fig.~\ref{fig:RSD8plots} are ranked
by their $D/T$ values, in face or edge-on, from the highest (left, upper panel) to lowest
(right, lower panel). The first four galaxies have 
$D/T>0.5$: they show relatively flat disks with
well defined spiral structures. Runs Sp2 and Sp6 have very thick disks and
no clear evidence of spiral structures. The two runs with the smallest $D/T$ values
are Sp4 and Sp5. The former, while in its edge-on projection shows
an oblate structure, in the face-on projection reveals well defined spiral structures. 
As will be shown below, this galaxy is assembled relatively late and 
had a recent merger and a burst of SF ($\sim 3-4$ Gyr ago);  
it seems that, as long as there is gas in the colliding galaxy, a 
thin disk with SF and spiral structure can be formed after the merger. 
The run Sp5, as discussed above, has a violent early and late-assembly 
history that has impeded it to form a disk component at any epoch; at $z=0$,
its kinematic disk-to-total ratio is only 0.1 (see Table~\ref{tab:sims}). 
This is compatible with what is seen in Figure~\ref{fig:RSD8plots}, 
that is an spheroidal structure.
Finally, note that in none of the runs the presence of
a clear bar at $z=0$ is observed. A similar result was reported by \citet{Aumer+2014}.
We will study this question in detail elsewhere.

In Figure~\ref{fig:SDuvSp7}, we show the U band face-on and edge-on projections
of run Sp7. The U band projection is color-coded in grey and its color bar goes from
5 to 6.7 in the log in units of $L_\odot/\kpc^2$.  This band samples
much better the young stellar population which, as can be appreciated
in the figure, resides in a thin disk and spiral arms.

To distinguish the spheroid from the stellar disk using kinematics,
it is usual to compute the probability distribution function of the 
stellar circularity $\epsilon$. This is defined as the ratio 
between the component of the
specific angular momentum perpendicular to the plane of the disk $j_z$ 
to the specific angular
momentum of a circular orbit at the stellar particle position $j_c$
\citep{Scannapieco+2009}. This latter is computed as $j_c = r V_c(r) = 
r \sqrt{\grav M(<r)/r}$, where $r$ is the distance of
the particle to the center and $M(<r)$ is the total mass inside $r$. 
A second definition is also often used in which the previous formula
for $j_c$ is substituted
by the maximum specific angular momentum the particle can have given
its binding energy $E$. In the second case $\epsilon$ is always $\le 1$. We
will use the first definition because in this case $j_c$ is 
straightforward to compute.

In Table~\ref{tab:sims}, we present two estimates of the kinematic disk-to-total
ratio of our runs: $D/T = (T-B)/T$, where $T$ is the
stellar mass in the galaxy and $B$ is the stellar mass of the spheroid,
defined as twice the stellar mass found in the counterrotating galaxy
stellar component \citep{Abadi+2003,Colin+2010}, and $f_{\rm disk} = 
M_s(\epsilon > 0.5)/\ms$. Although both estimates give similar results, we
see that for Sp1, Sp3 and, to a lesser degree for Sp2 and Sp7, the $D/T$ value is 
higher than the $f_{disk}$ one. This difference could be explained if there is 
a component with positive but low circularity values, that are thus not
considered in the $f_{\rm disk}$ counting. Figure~\ref{fig:epsall} show that this
is actually the case: there is an excess on the low-$\epsilon$ part of the 
symmetric disk distribution that peaks around $\epsilon = 0.9$. As anticipated
from the reading of the $D/T$ and $f_{\rm disk}$ columns in Table~\ref{tab:sims}
and from Fig. \ref{fig:RSD8plots},
runs Sp1, Sp3, Sp7 and Sp8, are clearly examples of disk galaxies; they 
have $D/T$ values close to those found by \citet{Marinacci+2014} for their
Aquarium halos (see their Figure 4, square brackets). For their two galaxies,
GA2 and AqC5, \citet{Murante+2015} found $D/T = 0.8$ and 0.77, respectively, similar
to the values of our Sp3 (0.81) and Sp1 (0.75) runs.
The runs Sp2 and Sp6 present a disk but it is not dominant, while 
runs Sp4 and Sp5 are by far spheroid-dominated galaxies.

\begin{figure}[htb!]
\plotone{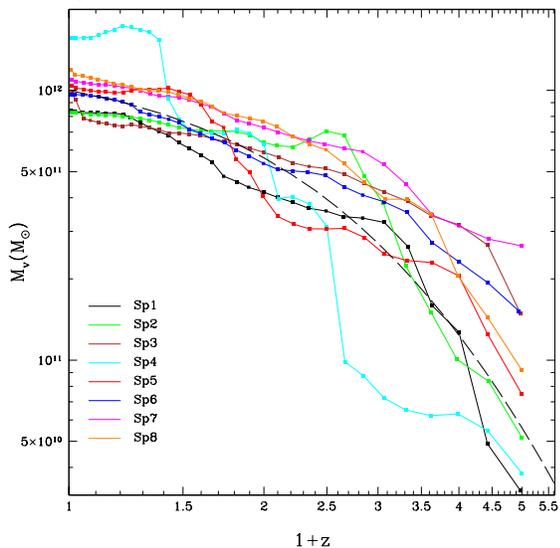} 
\caption[fig:MAHs]{Mass assembly histories of our suite of simulations. In the
Y axis is plotted the total mass: the sum of the masses of the DM, stellar, and gas 
components, all inside the current virial radius. We see that Sp4 (cyan) and Sp5 
(red) have had recent major mergers. The median assembly history reported in 
\citet{Rodriguez-Puebla+2016} for $\mv(z=0)\approx 10^{12}\ \msun$ halos is shown
as a black dashed line.}
\label{fig:MAHs}
\end{figure}

\subsection{Mass assembly histories} \label{sec:mah}

The halos were selected to be at $z = 0$ in relative isolation. Can they all be expected to harbor
a galaxy with a significant disk component? Not necessarily. It is known 
that the stellar contribution of this component is more strongly related rather 
to the halo mass aggregation histories (MAHs) \citep[see e.g.,][]{Scannapieco+2015}.
In Figure~\ref{fig:MAHs} we depict the (total) MAHs of our suite of 
simulations. For comparison, the median MAH reported in \citet{Rodriguez-Puebla+2016}
for $\mv(z=0)= 10^{12}\ \msun$ halos from several N-body large
cosmological simulations is also plotted (black dashed line).

Two of the runs, Sp4 (cyan) and Sp5 (red; see Table~\ref{tab:sims}), 
have a late violent assembly history, 
acquiring half of their present-day
masses at $\zhalf \simeq 0.48$ and 0.9, respectively. Sp5 actually increases its mass by about 
a factor of 2.5 from $z = 1$ to 0.4, and Sp4 have a major merger, quite visible in the
figure, around $z = 0.4$. The galaxies formed in these two halos are those with 
the lowest $D/T$ ratios, being completely spheroid-dominated. 
The run Sp2, which presents a significant spheroid component ($D/T=0.43$), had a significant 
merger at $z\sim 1.9$ and after that the mass increased very little.
Aside from these runs, the rest have more quiet late halo MAHs, 
forming half of its mass at $z > 1$; for example, $\zhalf \simeq 1.5$ for Sp8,
a disk-dominated galaxy with a very regular stellar mass assembly history.

In summary, we see that the halo MAH has a relevant effect on the galaxy morphology.
However, we have also a case where the formation of a prominent spheroid seems not
to be related to the MAH of the system, namely run Sp6 ($D/T= 0.30$). This run
presents a very regular halo MAH, actually without major mergers at any epoch but with
an early assembly of mass; its  SFR decayed long ago. Cases of spheroid formation
in simulated MW-sized galaxies, not related to the halo MAH, 
were reported and discussed previously,
for instance, in \citet{Sales+2012} and \citet{Aumer+2014}. These authors find that slowly rotating 
spheroids can form in systems where the angular momentum of the infalling gas is misaligned and
the gas is accreted directly through cold filaments; these events are likely to be 
major contributors to the central mass growth of the galaxy.

\begin{figure}[htb!]
\plotone{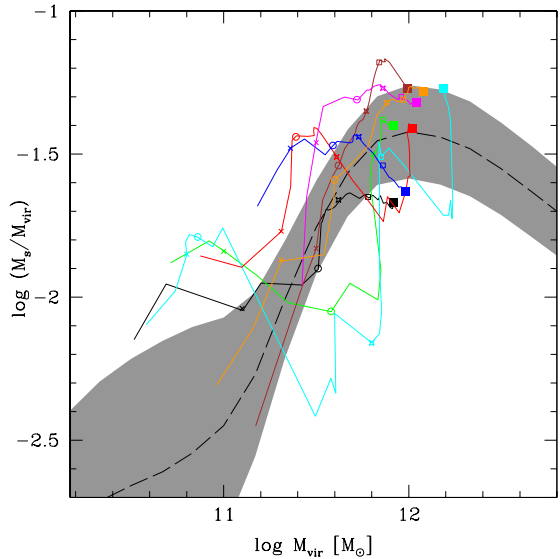} 
\caption[fig:msmh]{Evolution of the stellar-to-halo mass fraction plotted 
against the halo (total) mass for our suite of runs. 
With solid squares, empty squares, four pointed stars,
circles, and crosses are shown the values at $z = 0$, 0.54, 1.0, 2.07, and 3,
respectively. The trajectories begin at $z = 4$. The color-code is the same as in
previous figures. The gray shaded area and  the dashed line are the $1\sigma$ intrinsic scatter 
and the mean, respectively, inferred for $z\sim 0$ central galaxies in \citet{Aldo+2015}
It is interesting to see that the disk-dominated simulated galaxies move
in this plot nearly along the $z\sim 0$ stellar-to-halo mass relation.}
\label{fig:msmh}
\end{figure}

\begin{figure}[htb!]
\plotone{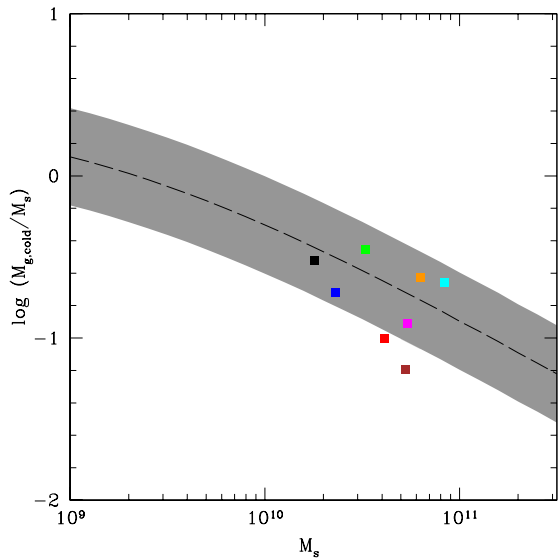} 
\caption[fig:Rgas]{The ratio of cold gas to 
stellar mass plotted as function of this latter. As in Figure~\ref{fig:TF},
with colored squares we show the predicted values by our set of runs. The
shaded area and the dashed line are the 1$\sigma$ intrinsic scatter
and its mean, respectively, from the observational compilation 
and homogenization by \citet{Calette+2016} for late-type galaxies.}
\label{fig:Rgas}
\end{figure}

\subsection{Stellar and Gas fractions}
\label{stellar-gas-fractions}

In Figure~\ref{fig:msmh}, the stellar-to-halo mass fraction, \ms/\mv, as
a function of \mv\ for our set of runs are plotted. The $\ms/\mv$ values at
$z = 0$ are shown as solid squares. The gray shaded area
($1\sigma$ error bars) is from the accurate semi-empirical determinations at $z\sim 0$
by \citet{Aldo+2015} for central galaxies (satellite galaxies present
a slightly different \ms--\mv\ relation).  Our simulated MW-sized galaxies
are consistent with the semi-empirical determinations. 

In Figure~\ref{fig:msmh},  we also depict with color lines the evolutionary tracks of the 
runs in the \ms/\mv--\mv\ plane.  We  show five redshifts along the tracks,
$z=0, 0.5, 1, 2, 3,$ and 4 (solid squares, empty squares, stars, circles, crosses,
and the end of the lines, respectively).  Interestingly enough, the most disky runs (Sp1, Sp3, Sp7, and Sp8)
evolve nearly along the semi-empirical $z=0$ \ms--\mv\ correlation, specially Sp3 and Sp8. 
The two most spheroid-dominated runs (Sp4, cyan line, and Sp5, red line)
present very irregular evolutionary tracks, likely due to the major mergers
they suffer. 
In general, the fact that the \ms/\mv\ evolutionary tracks of the
simulated galaxies do not deviate significantly from the $z=0$ \ms/\mv--\mv\ correlation,
at least up to $z\sim2$, would imply that this correlation does not evolve significantly in the
$10^{11}-10^{12}$ halo mass range.

In Figure~\ref{fig:Rgas}, the cold gas-to-stellar mass ratio, \mg/\ms, 
as a function of \ms\ are plotted for the eight runs. They are compared to the correlation
inferred from an extensive compilation and homogenization of observations for local late-type 
galaxies \citep[][gray band, corresponding to 1$\sigma$ intrinsic scatter]{Calette+2016}. 
We see that most runs are within the 1$\sigma$ intrinsic scatter of the observational 
correlation. Those that deviate most are run Sp5,
a spheroid-dominated galaxy with the lowest $D/T$ value of our sample (only 0.1),
and run Sp3, a disk galaxy with actually the highest $D/T$ value. As
we will see below, Sp3 had a second and long period of very active SF that
likely reduced the amount of cold gas at present-day, with respect to the other 
disk galaxies. The footprint of this second burst of SF 
and deficit in cold gas can also bee seen in its stellar-to-halo
mass fraction; the Sp3 galaxy has the second highest value of \ms/\mv\ among all runs
(see Figure~\ref{fig:msmh}).

An interesting question is what are the fractions of the different baryonic components
inside the virial radii of our MW-sized simulated galaxy/halo systems. In the case of stars and cold gas,
at $z=0$, they are located practically only within the central galaxies. The average and standard deviation of the 
\ms/\mv\ fraction normalized to the the universal baryonic mass fraction, $f_s= \ms/\mv/f_{\rm univ}$ 
($f_{\rm univ}\equiv \Omega_b/\Omega_m=0.15$ for our cosmology), from our eight simulations 
are $\langle f_s\rangle  =0.2775\pm0.1039$. For the cold gas, $\langle f_{\rm cold}\rangle  =0.0536\pm0.0275$, that is 
the cold gas mass is $\sim 1/5$ of the stellar mass, and
both account for approximately one third of the universal baryonic fraction inside the virial radius
of the halos. Where are the remaining baryons? 
A similar exercise was done but now for the
cool ($10^4<T/K\le10^5$), warm-hot ($10^5<T/K\le10^7$) and hot ($T>10^7 K$) gas. Most of the
mass of these gas components are located outside the galaxies, in the circum-galactic medium (CGM).
The average and standard deviations of the corresponding fractions (within 1\rv)
are: $\langle f_{\rm cool}\rangle  =0.0343\pm0.0283$, $\langle f_{\rm w-h}\rangle  =0.2425\pm0.0790$,
$\langle f_{\rm hot}\rangle  =2.7\times 10^{-6}\pm4.3\times 10^{-6}$. The sum of all of these baryonic
components does not account for the universal baryonic fraction. The missing mass is
$0.3925\pm0.1388$. Therefore, $\approx 25-53\%$ of the baryons in our MW-sized simulations
resides outside the corresponding virial radii.  

Recently, \citet[][]{Wang+2016} have reported the baryonic fractions mentioned above for their 
88 simulated galaxies, which cover a large range of masses (the NIHAO project). The average values reported
for their systems in the $0.3-3.5\times 10^{12}$ \msun\ halo mass range, roughly agree with those 
found here, though we predict a smaller fraction of cool gas ($\langle f_{\rm cool}\rangle  =0.034$ vs. 0.109) 
and a higher fraction of warm-hot gas ($\langle f_{\rm w-h}\rangle  =0.242$ vs. 0.167) than \citet[][]{Wang+2016}. 
These differences could partly be due to the fact that their averages include systems less massive 
than ours and, as they show, $f_{\rm cool}$ decreases and $f_{\rm w-h}$ increases for massive
systems.  Following \citet{Wang+2016}, we can compare the different gas fractions in the CGM with
recent observational constraints for the MW and $L^{*}$ galaxies using quasar absorption lines 
\citep[][see more references therein]{Werk+2014}. According to the latter authors, over $25\%$ of the
baryon budget is accounted for by cool, photo ionized gas in the CGM. This is $\sim 7\times$ higher
than the mean fraction measured in our simulations. On the other hand, the fraction of warm-hot gas in our
simulations is higher than conservative observational estimates \citep[e.g.,][$24\%$ vs. $5\%$]{Peeples+2014}
and well within the range of less conservative estimates. Finally, both our simulations and observations,
show that the fraction of hot gas is very small, though in our case is much smaller than estimates from $X-$ray
observations. As in  \citet{Wang+2016}, we conclude that while the total gas fraction measured in our
simulations ($\approx 35\%$ on average) seems to be consistent with the observational estimates
in $L^{*}$ galaxies, the mix of temperatures in the CGM strongly disagrees with current observational estimates. 


\begin{figure}[htb!]
\plotone{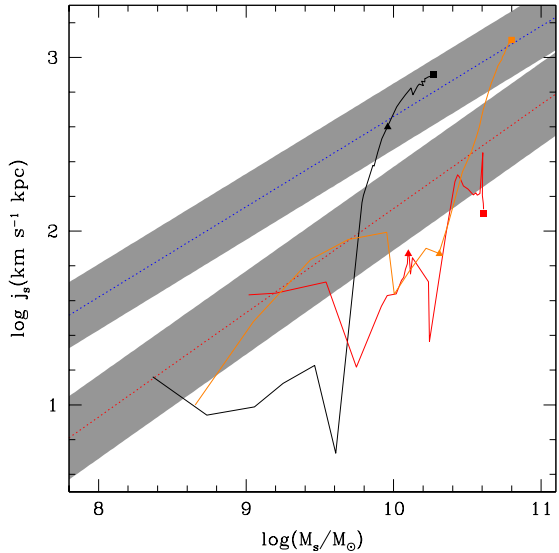} 
\caption[fig:jsevol]{Evolution of the specific stellar angular momentum $j_s$
plotted against the stellar mass for three of our runs: Sp1, Sp5,
and Sp8. Color code is as in Figure~\ref{fig:MAHs}. Straight dotted lines
are fits from \citet{RF2012}, red is for elliptical and blue for spiral galaxies.
The curves are plotted from $z = 4$ to present-day, this latter shown
with solid, colored squares. The points at $\sim \zhalf$ have highlighted
with solid triangles.}  
\label{fig:jsevol}
\end{figure}

\subsection{Evolution of angular momentum}

We compute the evolution of the stellar specific angular momentum, $j_s$, for our suite of
simulations and plot it as a function of the galaxy stellar 
mass in Figure~\ref{fig:jsevol}. For clarity, only the evolution of runs Sp1 and Sp8, 
two disk-dominated galaxies, and Sp5, a spheroid-dominated galaxy, are shown.
Lines start at $z = 4$, low $j_s$ values, and end 
at $z = 0$, shown in solid squares. We highlight with solid triangles the epochs at
which runs reach half of their stellar mass.
The straight dotted lines are observation-based fits 
from \citet{RF2012}; the blue one is for spirals 
while the red one is for ellipticals. The shaded area show the 1$\sigma$ scatter.

We found that all runs have low $j_s$ values at early times. This is in line
with the finding by \citet{AK2015}. They report similar values, although,
depending on the strength of the feedback, they fluctuate more or less.
Consistent with the $D/$T values shown in Table~\ref{tab:sims}, 
the Sp1 and Sp8 galaxies end up in the spiral region while the
Sp5 galaxy never leaves the elliptical zone and it actually ends up
below it. On the other hand, its late ($z  < \zhalf$) strongly fluctuating 
$j_s$ history is
consistent with the fact that more than half of its mass is assembled at $z < 1$.
We see that disky Sp1 and Sp8 runs (and also Sp3 and Sp7, not shown)
have a smooth late evolution in which $j_s$ increases all the time.
They differ each other on when this smooth period of angular momentum growth
begins; for example, for Sp1 it happens at $z \sim 1.8$, a little
later (0.7 Gyr) after the last major merger (LMM), while for Sp8 we can identify it
at $z \sim 1.3$, 2 Gyr after the LMM. 
The spheroid-dominated  Sp5 (as well as Sp4 and Sp6) runs have an
episodic evolution of $j_s$ but most of time their $j_s$ values are below or within
the region inferred from observations for local early-type galaxies.

\subsection{Star formation history}
\label{SFhistory}

\begin{figure}[htb!]
\plotone{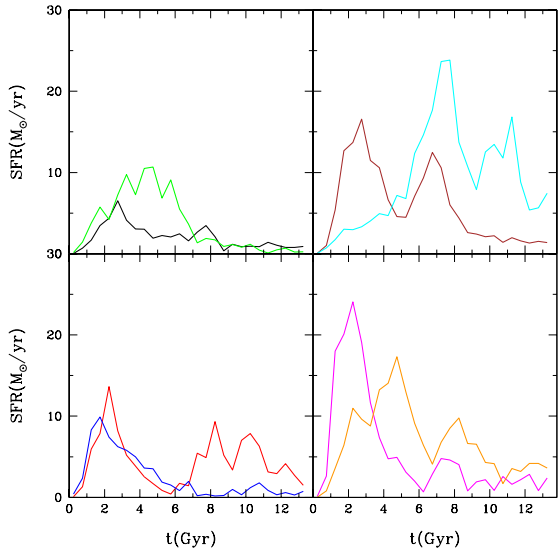} 
\caption[fig:sfh]{Star formation rate, measured as a stellar age histogram, as a function of cosmic time,
in bins of 0.5 Gyr width, for all the eight runs. We place in each panel two runs and they are color-coded as
in Figure~\ref{fig:MAHs}; for example, Sp1 (black line) and Sp2 (green line) are in the top left panel, and 
Sp7 (magenta line) and Sp8 (orange line) are in the bottom right one. All runs show a prominent first peak,
related to the violent phase of the assembly history. The disk-dominated runs Sp1, Sp3, Sp7 and Sp8 also present 
a second peak, produced well after the last major merger, during the smooth growth of the angular
momentum.}  
\label{fig:sfh}
\end{figure}

Aside from Sp2, all runs, even those with a massive spheroid component, show
present-day SFRs (see Table~\ref{tab:sims}) that are comparable to those 
of late-type spiral galaxies at $z = 0$ \citep[c.f.][]{Schiminovich+2007}. 
In particular, the most rotational supported disks have values close or even 
higher than those reported for the 
MW \citep[e.g.,][]{RW2010}, consistent with the gas fractions 
predicted for these runs. In Figure~\ref{fig:sfh} we show the ``archaeological''
SF history (SFH) for our set of runs. This is determined from the snapshot at 
$z = 0$ by simply counting the mass in stellar particles in a given age
(time) bin and divided it by the width of the bin. We divide the time
since Big Bang in bins of 0.5 Gyr. Thus, in each panel, in the left (right) 
we have the oldest (youngest) stellar particles.

In agreement with the many SFHs shown in works on the subject
\citep[e.g.,][]{Governato+2004, Guedes+2011, HB2012, Murante+2015}, we also see
in all of our runs (but Sp4) the characteristic early peak of SF, 
registered in the violent phase of halo mass assembly. The strength
and duration of the ``burst'' vary with the run. Somehow unexpected,
in the most disky runs (with the highest kinematic $D/T$ values), one sees
a second peak of SF located in time well after the LMM,
during the smooth growth phase of the angular momentum. 
Regarding 
the most spheroid dominated galaxies, Sp4 (cyan) and Sp5 (red), they show
relatively strong SF activity in the last few Gyr as the
result of their late mergers. Thus, the origin of spheroid-dominated MW-sized
galaxies in a relatively isolated environment 
can be related in some cases
to late mergers with significant fractions of gas and ulterior bursts of (recent) SF. 
Therefore, one expects that a fraction of MW-sized early-type galaxies in 
the field have a different structure and stellar population composition
as compared to cluster early-type galaxies of similar masses.


\begin{figure}[htb!]
\plotone{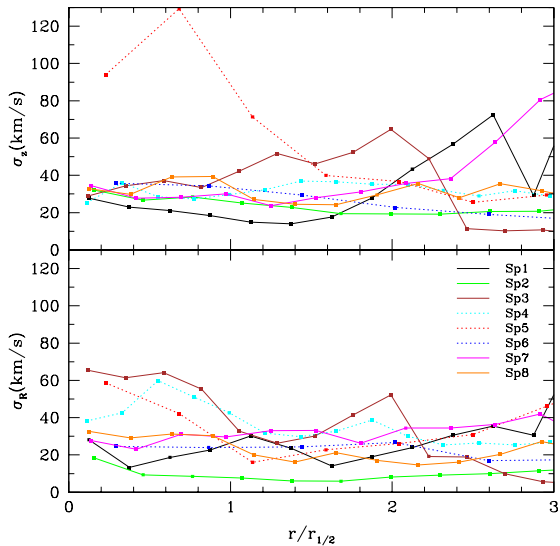} 
\caption[vel-disp]{Azimutally-averaged velocity dispersions in the vertical (upper panel) and radial (lower panel)
directions for the runs with $D/T>0.4$ at $z=0$, Sp1, Sp2, Sp3, Sp7, and Sp8 (solid lines) and those with
 $D/T\le0.4$, Sp4, Sp5, and Sp6 (dotted lines).
}
\label{vel-disp}
\end{figure}

\subsection{Gas velocity dispersions}

A relevant ISM property is the cold gas velocity dispersion. The large-scale SF of disk 
galaxies could proceed by a self-regulation mechanism in the ISM driven by the balance
between the injected (by SNe) and dissipated kinetic energy in the ISM 
\citep[e.g.,][]{Firmani+1992,Avila-Reese+2000,Tamburro+2009,Klessen+2010}. 
In this context, the cold gas velocity dispersion is directly related to the SN-driven feedback. 
We will discuss this self-regulation model in the light of our simulations in subsection \ref{selfregulation}. 

In Fig. \ref{vel-disp}, we present the gas surface density-weighted velocity dispersion profiles of
the simulated galaxies at $z=0$ for both the radial and vertical components, $\sigma_r$ and $\sigma_z$, respectively. 
Only gas cells with $T<10^4$ K and $|z| \le  1$ kpc, where $z$ in this case is the $z$ coordinate,
perpendicular to the plane, are taken into account. 
The runs with $D/T>0.4$, Sp1, Sp2, Sp3, Sp7, and Sp8, are shown with solid lines while those
with $D/T\le0.4$, Sp4, Sp5, and Sp6 are shown with dotted lines. 

\begin{figure}[htb!]
\plotone{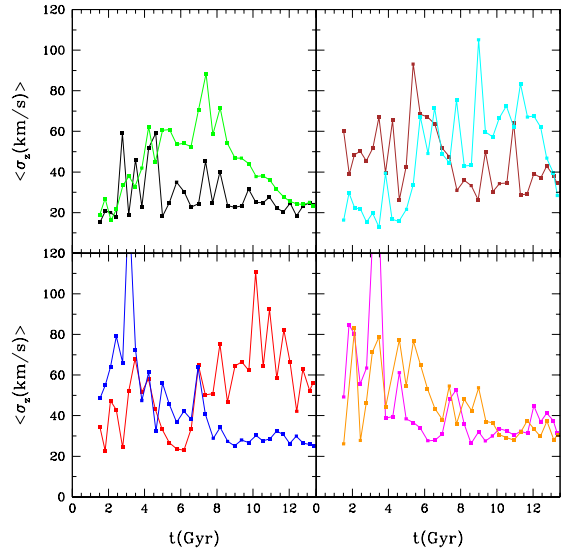} 
\caption[vel-disp]{Evolution of the mean surface density-weighted vertical velocity dispersions of the
eight simulated galaxies. 
}
\label{vel-disp-evol}
\end{figure}

The gas velocity dispersion is moderately anisotropic, more for the spheroid-dominated galaxies. 
The $\sigma_z$ profiles are nearly flat with values mostly around 25-35 \kms\ up to $\sim 1.5\rhalf$,
except for Sp5. The latter galaxy had recent mergers and strong bursts of SF; therefore, the small fraction
of remaining cold gas is still strongly turbulent. In several cases, $\sigma_z$ increases with radius
in the outermost regions, above 1--1.5 $\rhalf$.
The radial velocity dispersion profiles,  $\sigma_r$($r$), are also nearly flat for the disk-dominated
galaxies (with values around 15-30 \kms) and with higher values in the inner regions 
for the Sp5 and Sp6 runs, which are spheroid-dominated galaxies that suffered relatively recent mergers.  
The Sp3 run, while disk-dominated, presents large values of $\sigma_r$ in the inner disk and 
increasing values of  $\sigma_z$ in the outer disk; the gas of this galaxy is being perturbed by a
relatively massive ($M_s \sim 10^{10}\ \msun$) satellite that is inside \rv\ at $z = 0$.

The observational measure of the cold (HI and/or CO) gas velocity dispersion profiles
in galaxies is not an easy task. For data cubes, the gas velocity dispersion 
is calculated from the  second-moment map. For example, in \citet{Tamburro+2009} the data cubes of 11 galaxies from 
the ``The HI Nearby Galaxy Survey'' \citep[THINGS;][]{Walter+2008} are used to calculate the line-of-sight
(intensity-weighted) velocity dispersion at each pixel.
In \citet[][see also \citealp{Caldu-Primo+2013}]{Mogotsi+2016}, 
the line-of-sight (natural-weighted) velocity dispersion is calculated from Gaussian fits to the line profiles. 
These authors studied both the THINGS HI and the ``HERA CO Line Extragalactic Survey'' (HERACLES) 
CO line profiles of 13 galaxies; they present results also using the second moments of the HI profiles. 

For spiral MW-sized galaxies, the mentioned above observational results show line-of-sight velocity dispersions that
mostly decrease with radius. For most cases, the
velocity dispersion profiles flatten in the outer regions, attaining values around $10$ \kms. 
The mean/median values of the line-of-sight velocity dispersion of observed galaxies range
from $\approx 8$ to 20 \kms, which are up to $\sim 1.5$ times lower than the mean surface
density-weighted vertical and radial velocity dispersions of our simulated galaxies, $\langle  \sigma_z\rangle  $
and $\langle  \sigma_r\rangle  $, respectively.  
It is difficult to conclude whether there is a disagreement or not between simulations and observations, because of
the large differences in the way of measuring the velocity dispersions among different observational studies and 
the likely biased way simulations of comparing simulations with observations. Within this level of uncertainty, we 
can just say that the velocity dispersion profiles in our simulations are not in strong conflict
with those inferred from different observational studies. This implies that we are in the limit of the stellar 
feedback efficiency because the cold gas velocity dispersion (and gas disk thickness) significantly
increase with this efficiency \citep[see][]{Marasco+2015}.

In Fig. \ref{vel-disp-evol}, the evolution of $\langle  \sigma_z\rangle$ is  shown for the eight MW-sized runs. 
By comparing this plot with figure \ref{fig:sfh}, we see a
trend of the $\langle  \sigma_z\rangle  $ history to roughly follow the global SF history. In general, the velocity dispersions 
of the simulated galaxies are larger at higher redshifts, when the galaxies are having
frequent mergers and strong bursts of SF. Recent observational inferences of CO velocity
dispersions in high redshift galaxies, show that they are indeed characterized by clumpy gaseous disks
much more turbulent than the local ones. The measured velocity dispersions of star forming galaxies
at $z\sim$1--3 are close to their rotation velocities, unlike in local galaxies, where the former velocities
are much lower than the latter ones \citep[e.g.,][]{Law+2007,Law+2009, Forster-Schreiber+2009,Epinat+2012,Lehnert+2013}. 

\section{Discussion} 
\label{sec:discussion}

\subsection{Resolution issue}

In \citet{Santi+2016}, they compared their results from
the high resolution simulation G.321 ($7 \times 10^6$ DM particles
inside \rv\ and 109 pc of resolution) with 
the ones from G.321$_{lr}$, a simulation with twice less resolution and about
eight times less DM particles, and found good convergence except for the SFR and the
amount of cold gas
(the comparison is only made at $z = 0$). 
For the former, they report 0.27 \msun/yr for G.321 against 0.16 for G.321$_{lr}$,
a factor of 1.7 higher for the high resolution run. For the latter, a higher difference is found,
$9.3 \times 10^9 \msun$ for G.321 against $1.4 \times 10^9$ for G.321$_{lr}$. Guided by these results 
and the resolution study of \citet{Scannapieco+2012}, we run again the Sp7 galaxy but with one less
level and eight times less DM particles. For this low resolution simulation, the values of the total, 
stellar, and cold gas mass inside the galaxy are: $1.06 \times 10^{12}$, $5.05 \times 10^{10}$, and $3.76 \times 10^9\ \msun$,
respectively. Such values differ from the corresponding ones in the higher resolution simulation by
2.7\%, 3.6\%, and 41\%, respectively (see Table 1). Moreover, the $V_{\rm max}$ value differ by less than 1\%. Overall, 
we see that for global and cumulative quantities, like the stellar mass, good
convergence is achieved. On the other hand, we have found in the low 
resolution simulation a lower present-day SFR and lower specific angular momentum. The
values are: (SFR/\msun\ yr$^{-1},j_s$/km s$^{-1}$ kpc)=(0.7,605). These are to be compared with (3.2,957),
found for the higher resolution counterpart. That the $j_s$ value is lower in the low resolution run 
is not surprising because it is long known that $j_s$ does depend on resolution \citep{Governato+2004}.
Aside from the two-body heating and the dynamical friction effect of the halo on the 
accreting lumpy substructure,
there is also the torque exerted by the halo on an asymmetric, not well resolved, disk
\citep{Mayer+2008}. It seems, however, that resolution through  artificial loss of angular 
momentum is also the culprit of the lower SFRs and cold gas mass values reported above. 

Have we reached
convergence given our SF and feedback scheme and the corresponding chosen parameters?
This is a question that could be addressed by running simulations with even more 
resolution. Unfortunately, this can not be performed with the present set of simulations, so we
can only speculate. On the other hand, the experiment by \citet{Santi+2016} seems to suggest 
that $\sim\ 10^6$ DM particles and $\sim\ 220$ pc resolution are not enough, at
least as the SFR and amount of cold gas is concerned.\footnote{In the resolution
study of \citet{Marinacci+2014}, they found, in the Aquarius C series of simulations, 
different SFRs and gas masses values but not as much as those found here or
in \citet{Santi+2016}. Interestingly, they found SFR and gas mass actually decrease, by 
a factor of 1.9 and 2.0, respectively, from low (Aq-C$_5$) to high (Aq-C$_4$)
resolution (Marinacci, private communication).}
It is not clear, on the other hand, if the present stellar feedback scheme, considered in this and 
other related papers (see, for example, \citet{Santi+2016}), can be applied to
much smaller scales ($\sim 50$ pc or below) because at these scales physical processes, 
such us the gas heating by ionizing radiation \citep[e.g.,][]{Hopkins+2014, Trujillo-Gomez+2015}
should be included. It is also possible that, because our stellar feedback is deterministic, 
convergence (for a given \esf\ and \nsf) may not be achieved since twice smaller cells 
(two times better resolution) would imply eight times less massive SPs on average. Thus,
twice higher resolution means that the increase of temperature (to several $10^7 \Ke$) 
happens in only 1/8 of the cell of the less resolved simulation. This process could
be non-linear and therefore convergence may not be guaranteed.

\subsection{Star formation and stellar feedback}
\label{SFandFeedback}

The present version of the (deterministic) SF and stellar feedback recipes 
have been used since the paper by \citet{Colin+2010} 
\citep[see also, e.g.,][]{Avila-Reese+2011,Gonzalez+2014}. In particular, the paper by \citet{Colin+2010} focused
on the impact of varying the SF and stellar feedback parameters on the properties of the simulated
low-mass galaxies. For instance, it was seen that a lower \nsf\ produced a larger \rhalf\ and a less
peaked circular velocity. Moreover, because the thermal SNe energy is dumped 
into the target gas cell instantaneously, the lower \esf\ the less efficient the feedback is. Unfortunately, because it is not implemented
in the present version of the code, we can only speculate about what would happen if part of this energy 
were kinetic. It is interesting, on the other hand, to see that while this prescription produces 
a stellar-to-halo mass ratio of MW-sized galaxies that agrees with semi-empirical determinations, 
this is not
the case for our simulated low-mass galaxies \citep[see figure 6 of][]{Avila-Reese+2011}.

\subsection{``Explosive'' feedback}
\label{feedback}

In {\it every} cool and dense gas cell we
let the code convert 65\% of its mass into stars. This high fraction is really needed
to overcome the overcooling problem. This spatial and time located injection of energy 
increase the temperature of the gas to several $10^7$ K degrees. 
This temperature is obtained by matching the expression for the internal
energy of an ideal monoatomic gas with the energy injected by SNe  
associated with the stellar particle; that is, $T$ is given
by
\begin{equation}
T = \frac{2}{3} \left( \frac{1}{Nk} \right) 2 \times 10^{51} N_{SNII},
\end{equation}
where $N_{SNII}$ is the number of supernovae, which depends on the
shape of the IMF and the mass of the stellar particle, and $N$
is the number of hydrogen atoms inside the cell where the SP was
born. This latter is given by $n_H(1-\esf)\Delta x^3$, 
where $\Delta x$ is the side of the
cell, $n_H$ is the hydrogen number density which is
greater but close to \nsf, and $\esf = 0.65$. We notice that this
tempearture does not depend on the size of the cell because  
$N$ and $N_{SNII}$ both depend on $\Delta x^3$, the latter 
through the mass of the stellar particle.
Since the gas density
in the cell at the moment of SF is around \nsf\ (in our case 1 hydrogen
atom per cubic centimeter), the gas pressure right after the SF event is huge, $10^7\
\Ke \cmtres$, clearly enough to overcome the external pressure and the 
negative ``pressure'' due to gravity \citep{CK2009, Trujillo-Gomez+2015}. At these
temperatures the crossing time is much smaller than the cooling time \citep{DS2012}
and thus, delaying the cooling after SF, to avoid the gas from radiating away most of 
its heat, becomes irrelevant\footnote{Yet, as the cooling is
turn off where the young stellar particle {\it is} and the particle can leave
the cell (where it was born) during the time the cooling is off (40 Myr), a non-neglible 
effect on the properties of galaxies could still be expected.}. Yet, as the simulations in \citet{Santi+2016} were
run assuming this delay in the cooling our simulations also consider this 
assumption. 

\subsection{Dark and baryonic mass assembly of MW-sized galaxies}

Galaxies formed in halos of present-day masses $\mv\approx 10^{12}$ \msun\ are in the
peak of the \ms/\mv\ correlation (see e.g., the gray area in Fig. \ref{fig:msmh}); that is, galaxies of these masses
were the most efficient in forming stars within their halos. As discussed in Subsection \ref{stellar-gas-fractions}, our 
simulations are in rough agreement with the semi-empirical inferences of the \ms/\mv\ correlation, and the most 
disk-dominated ones (Sp1,Sp3, Sp7, and Sp8) evolve closely along the $z=0$ \ms/\mv--\mv\ correlation.

\begin{figure}[htb!]
\plotone{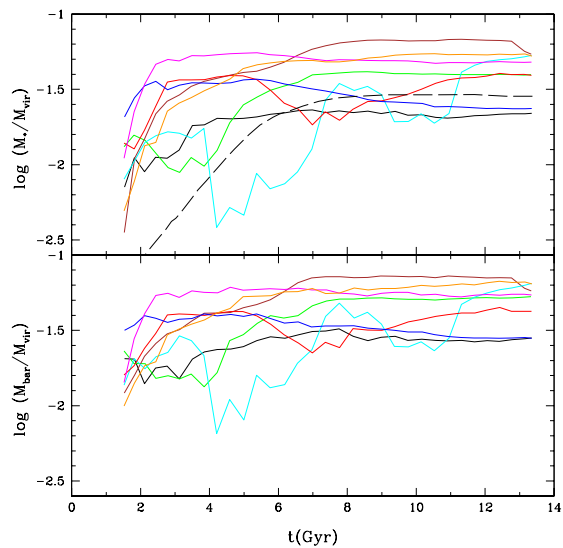} 
\caption[fig:MsMbar]{Evolution of the \ms/\mv\ (top panel) and $M_b/\mv$ (bottom panel)
fractions for the eight runs. The color code is the same as in Fig. \ref{fig:vc}. 
The dashed line in the top panel is the average track for $z=0$ $\mv=10^{12}$ \msun\
halos obtained from the look-back time semi-empirical inferences of \citet[][]{Behroozi+2013}.}
\label{fig:MsMbar}
\end{figure}

Is the \ms/\mv\ evolution of our MW-sized galaxies consistent
with look-back time semi-empirical inferences?  In Fig. \ref{fig:MsMbar} (upper panel) we compare
the \ms/\mv\ tracks (solid lines) from the simulated galaxies with the average track
for $\mv=10^{12}$ \msun\ halos according to the inferences by 
\citet[][dashed line]{Behroozi+2013}. Five of the simulated galaxies (those with the 
largest $D/T$ ratios) present a nearly constant value of the \ms/\mv\ fraction since 
$\sim$6.5--10 Gyr ago, and then a fast decrease at earlier epochs.\footnote{
The two runs that end up with the lowest $D/T$ ratios, Sp4 and Sp5, 
show \ms/\mv\ tracks with periods of increase and decrease, corresponding to the
epochs of major mergers. However, along their evolution, the \ms/\mv\ values of these runs 
do not shift significantly from the values of the other runs. }
A similar behavior is seen for the average track inferred from the semi-empirical study.
It is interesting to see that for MW-sized galaxies, both semi-empirical inferences
and the simulations show that their stellar masses grow roughly as their halo masses
during the last 6.5--10 Gyr (since $z\approx$ 0.7--2); at the earliest epochs, these galaxies
were inefficient in forming stars  (their \ms/\mv\ fractions were
below 1--3$\times 10^{-3}$) and then, at ages around 2--6 Gyr, when the initial burst of SF happens,
the stellar mass grew up at a rate much faster than the halo mass, to finally enter
in the quiescent regime in which $\ms(z)\propto \mv(z)$ roughly. 

Aside from the fact that uncertainties in the determination of the \ms/\mv\ correlation at high redshifts are
large, the inferences by \citet[][]{Behroozi+2013} shown in Fig. \ref{fig:MsMbar} suggest that 
galaxies formed in halos of present-day masses around $10^{12}$ \msun\ assembled their stars
later on average than our simulations. It is possible that at the early stages of
galaxy formation, when the host halos were much less massive than today,
the effects of early or preventive stellar feedback (see the Introduction for references) reduce
significantly the SF. Moreover, we notice that the AGN-driven feedback
is not yet implemented in the code. So, it could be that  in our simple scheme
too many stars are being formed in the earliest epochs. In any case, given the good agreement
between observations and our simulated galaxies, there is not much room for
other feedback effects, at least at the MW-sized scales studied here.

It is interesting that the epochs at which the simulated galaxies reach the quiescent regime
correspond roughly to the epochs where the halos attained a mass of $\approx 5\times 10^{11}$ \msun. 
Several semi-empirical studies have suggested that halos in the mass range $\mv\approx 0.5-1\times 10^{12}$ \msun\ 
are those with the maximum SF efficiency at any epoch \citep[e.g.][]{Behroozi+2013b,Lu+2015}. 
This can be explained as follows. The stellar feedback is able to blowout the gas from the galaxies with a 
efficiency that decreases as the gravitational potential gets stronger. Therefore, the more massive the galaxy/halo 
system is, the less gas is ejected due to stellar feedback. However, for halos above $\sim 10^{12}$ \msun, 
other astrophysical processes start to play an
important role, making the efficiency of star formation to decrease with mass. 
These processes are the long radiative cooling time of the gas, shock heated 
during the virialization of massive halos \citep[see e.g.][]{Dutton+2010}, and 
the feedback of powerful AGNs, which 
appear at some evolutionary stages of massive galaxies \citep[see for a review e.g.,][]{Somerville-Dave2014}.
The consequence of this is that galaxies evolving in halos with current masses 
of $\sim 0.5-1\times 10^{12}$ \msun\ are expected  to be 
the most efficient in capturing, keeping, and transforming gas into stars. For this reason, the stellar mass growth
of these galaxies is expected also to be the less detached from the corresponding cosmological halo MAH.
Nevertheless, even these systems with the highest efficiency of stellar mass growth,
have \ms/\mv\ values much lower than the universal baryonic fraction, $f_{\rm univ}\equiv\Omega_b/\Omega_m=0.15$ 
for our cosmology. 

We can measure the baryonic mass of our simulated galaxies, $M_b=\ms + \mg$, and calculate
the galaxy baryonic fractions, $M_b/\mv$. The evolution of these fractions are plotted in the lower panel 
of Fig. \ref{fig:MsMbar}. While the values of 
$M_b/\mv$ are larger than those of \ms/\mv, the former are still lower than $f_{\rm univ} $, 
at least by a factor of $\sim 3$ (see also Subsection \ref{stellar-gas-fractions}).
The $M_b/\mv$ evolutionary tracks are similar to the \ms/\mv\ ones; nonetheless, at earlier epochs,
there are somee differences which are related to the fact that the 
gas-to-stellar mass ratios increase with redshift. 

As shown in Subsection \ref{stellar-gas-fractions}, at $z=0$ a large fraction of baryons in the simulations 
are located actually outside the central galaxies, in the gaseous circumgalactic medium, in different phases 
\citep[see also][]{Sales+2012,deRossi+2013,Santi+2016}. We have found that on average 
$3.4\%$, 28.3\%, and much lower than 1\% of the universal baryon fraction is contained in cool, warm-hot, and hot gas 
in the halo (within \rv), respectively (for comparison, the mass fraction in stars in the 
simulated galaxies is on average $\approx 28\%$). Stellar feedback is  
presumably the culprit for keeping the gas in the halo in these phases. However, even taking 
into account these contributions, the baryonic budget within the virial radius of our simulations present 
an average deficit of $\approx 40\%$ with respect to $f_{\rm univ} $. Therefore, yet a significant 
fraction of baryons should be in form of gas at distances greater than virial radii.
A detailed study of the origin and spatial distribution of the CGM in general will be presented elsewhere.

\subsection{What does drive the gas velocity dispersion in galaxies?}
\label{selfregulation}

Since MW-sized disk galaxies suffer less the influence of large-scale
SN- and AGN-driven galactic winds, they are optimal to study whether the SF proceeds
locally in a self-regulated way and whether the turbulent motions in the disks are driven by
SN feedback or some other mechanisms. While the SF in the disk 
can be explained by the Toomre \citep{Toomre1964} or generalized Toomre \citep{Romeo+2011} criteria, 
the physical conditions implied in these criteria are related namely to the SF and 
its feedback, as well as to properties of the turbulent ISM and the dynamics of 
the disk. In this context, one expects a self-regulated SF process controlled by ISM properties 
like the turbulence decay time, $t_{\rm d}$, or its porosity due to SN-remnant heated
gas, as well as by the gas infall rate \citep[e.g.,][]{Firmani+1992, Silk1997,Klessen+2010}. 

A simple scheme of local self-regulated SF is the one based on the vertical balance between
the rate of injected energy versus the dissipated kinetic one, which determines the gas disk height 
\citep{Firmani+1992,Firmani+1994,Avila-Reese+2000}:
\begin{equation}
\gammaE  \epsSN \dot{\Sigma}_{*}(r) = \frac{\Sigma_g(r)  \sigma_z^2(r)}{2 t_{\rm d}},
\label{balance}
\end{equation}
where $\dot{\Sigma}_{*}$ and $\Sigma_g$ are the surface SF rate and neutral gas density, 
$\gammaE$ is the SN + Stellar Winds (from stars $>8\msun$) energy fraction that ends 
as kinetic energy in the ISM,  and 
$\epsSN=E_{\rm SN+Wind} \etaE$ is the SN + Stellar Wind energy per unit 
of solar mass formed. In this equation, it is implicitly assumed that the gas velocity dispersion is driven
only by SN + Wind feedback. We calculate $\sigma_{z}$($r$) from our simulation results by using the
measured radial profiles $\dot{\Sigma}_{*}$($r$) and $\Sigma_g$($r$) (for the neutral, $T<10^4$ K, gas),
and compare the obtained $\sigma_z$($r$) profiles with those measured. For this, we use the same value of  
\epsSN\ as in the simulations ($E_{\rm SN+Wind}= 2 \times 10^{51}$ erg and $\etaE =7.5\times 10^{-3}$ $\msun^{-1}$ 
for a Miller-Scalo IMF). For $\gammaE$, the analytical solution of the SN-remnant evolution shows that 
it is equal to $\sigma_z/v_w$, where the terminal wind velocity is $v_w=340$ \kms\ \citep{Spitzer1978}. Numerical
simulations of SN-remnant evolution have shown that $\gammaE \approx 0.1$ \citep{Thornton+1998}, 
close to the analytical solution when the ISM velocity dispersion is 25-40 \kms, as in our simulations at $z\sim 0$.
Observational inferences agree also with vales of $\gammaE \approx 0.1$ \citep{Tamburro+2009}.
We will use $\gammaE=\sigma_z/v_w$. For the turbulence decay, several numerical studies have 
demonstrated that supersonic turbulence dissipates rapidly, in a timescale roughly equivalent to the crossing 
time for the driving scale at the turbulence velocity \citep[e.g.,][]{MacLow1999,Stone+1999,Avila-Reese+2001}.
The latter authors have found values of $t_d\approx 15-20$ Myr for ISM properties typical of the MW. 
Here, we will assume a constant $t_d$ equal to $15$ Myr.

\begin{figure}[htb!]
\plotone{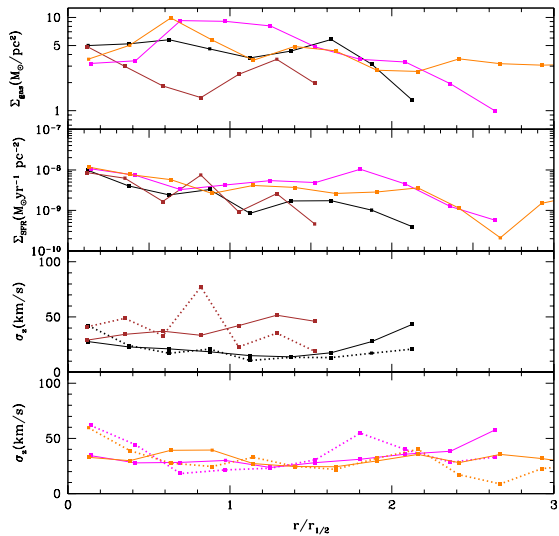} 
\caption[profiles]{{\it Upper panels:} Gas surface density and surface SF rate profiles at $z\sim 0$ for the 
disk-dominated runs (Sp1, Sp3, Sp7, and Sp8).  The color code is the same as in Fig. \ref{fig:vc}. 
{\it Lower panels:} Gas vertical velocity dispersion profiles as measured for the
same runs shown in the upper panels (solid lines) compared to the profiles calculated
from $\Sigma_g$ and $\dot{\Sigma}_{*}$ profiles according to Equation \ref{balance} (dotted lines). 
}
\label{profiles}
\end{figure}

In Fig. \ref{profiles}, we present the $z=0$ $\Sigma_g$($r$), $\dot{\Sigma}_{*}$($r$), and $\sigma_z$($r$)
radial profiles of our disk dominated galaxies (Sp1, Sp3, Sp7, and Sp8; solid color lines).
The $\sigma_z$($r$) profiles are presented in the two lower panels and they are compared
with the corresponding velocity dispersion profiles calculated using Equation \ref{balance} (color dashed lines). 
In general, the simulated galaxies roughly follow the azimutally-averaged energy balance condition given by 
Equation (\ref{balance}) along the disk, showing that the (vertical) turbulent motions are mainly driven by the 
SN feedback and that SF proceeds in the disks at $z\sim 0$ in a  nearly self-regulated fashion.  

In more detail, in three of the runs (Sp1, Sp7, and Sp8) the measured velocity dispersion is slightly larger than 
the calculated one at large radii, where the SF rate is already very low. This means that the gas motions 
at these radii are driven by other mechanisms outside SN feedback; for example, by the kinetic energy of the 
accreting gas from the galaxy surrounding environment \citep[][]{Santillan+2007,Klessen+2010}.
For three runs (Sp1, Sp3, and Sp8), the calculated values of $\sigma_z$ tend to be higher than the measured ones
in the central regions (for Sp3 this extends almost to $\rhalf$). This could be because the turbulence
decay time is actually not constant as assumed here but it may depends on the disk dynamics. For example,
the approach of turbulent gas cell collisions given in \citet[][]{Firmani+1992,Firmani+1994,Silk1997}, 
implies a $t_d$ that decreases toward the center, for a typical galaxy rotation curve, specially where 
the rotation curve rises steeply. Therefore, $t_d$ could decrease in the innermost regions, implying
lower values for the calculated $\sigma_z$ at small radii. We will study this question in detail elsewhere.

The results presented in Figure \ref{profiles} are at $z\sim 0$; i.e., when our simulated galaxies are evolving 
secularly.  As shown in Figure \ref{vel-disp-evol}, the gas velocity dispersion can attain high values in the past,
in most cases after the galaxy have had strong bursts of SF.  However, it is not clear whether such high
velocity dispersions can be maintained by SN feedback. At these epochs, the evolution of the disks
is likely happening in a regime dominated by large-scale gravitational instabilities and strong gas accretion,
as some observational inferences suggest \citep{Krumholz+2016}.
These processes have been found, in previous numerical simulations, 
to be the main drivers of the high velocity dispersions in high-redshift disks
\citep[see e.g.,][]{Agertz+2009,Bournaud+2009,Ceverino+2010,Goldbaum+2015}. 
This question will be analyzed in more detail elsewhere.

\section{Conclusions}
\label{conclusions}

A suite of eight MW-sized simulated galaxies (virial masses around $10^{12}\ \msun$ at $z=0$)
were presented. The hydrodynamics + N-body ART code and a relatively simple
subgrid scheme were used to resimulate with high resolution these galaxies. 
The regions, where the galaxies are located, are resolved mostly with cells 
of 136 pc per side and their halos contain around one million DM particles.
The resimulated regions were chosen from a low-resolution N-body only DM simulation of a box 
$50\ \mpch$ size of side. These regions target present-day halos in relatively 
isolated environments. The overdensities, concentrations, and spin parameters of the runs cover a wide
range of values. Our main results and conclusions are as follow:

$\bullet$ The {\it simple} but {\it effective} subgrid scheme used in our simulations works very well for producing disk
galaxies at the MW scale in rough agreement with observations. They have nearly flat circular velocity curves and
maximum circular velocities in agreement or slightly higher than the observed Tully-Fisher relation. Other predicted
correlations as the \re--\ms, \ms/\mg--\ms\ and \ms/\mv--\mv\ ones are consistent with observations for
disk galaxies. In fact, four of the simulated galaxies end up as disk dominated, two have a significant disk 
but it does not dominate, and two more end up as spheroids, with a small disk component.
The latter ones, as expected, moderately deviate from some of the mentioned correlations. 
Our subgrid scheme is based on a deterministic prescription for forming stars
in every cool and dense gas cell with a high efficiency, and in a ``explosive'' (instantaneous energy release)
stellar thermal feedback recipe. The parameters of the SF+feedback scheme for our given resolution,
were fixed in order to attain a high gas pressure, above $10^7$ Kcm$^{-3}$, in the cells where young
($< 40$ Myr) stellar particles reside. In this way, (1) the external pressure and the negative ``pressure'' 
due to gravity are surpassed allowing the gas to expand, and (2) the temperature grows 
high enough as to obtain crossing times in the cell that result much smaller than the cooling time. 
The chosen parameters are $n_{\rm SF} =1$ cm$^{-3}$, $T_{\rm SF}=9000$ K, 
$\epsilon=65\%$, $E_{\rm SN+Wind}=2\times 10^{51}$ erg.

$\bullet$ The disk-dominated runs are associated to halos with roughly regular MAHs (no major mergers,
at least since $z\sim1$), and they have a late quiescent stellar mass growth nearly proportional to the halo 
mass growth ($\ms/\mv\approx$const. since the last 6.5--10 Gyr). On the
contrary, the two most spheroid-dominated runs are
associated to halos that suffered (late) major mergers. However, a galaxy with
a prominent spheroid is also formed in a run
with a very regular (no major mergers) halo MAH; in this run, most stars are assembled in an early burst.
We conclude that the halo mass aggregation and merger histories play an important role in the
final galaxy morphology of our MW-sized systems, but other effects can also be 
at work in some cases.

$\bullet$ The disk-dominated runs present a gently and significant increase with time of the stellar 
specific angular momentum, $j_s$, attaining values at $z=0$ as measured in observed local
late-type galaxies. On the contrary, the spheroid-dominated runs present an episodic evolution 
of $j_s$, ending with low values, as measured in the local early-type galaxies. 
Moreover, the SFR histories of seven of the runs present a strong initial burst at the age
of $2-4$ Gyr and then a decline with some eventual bursts. The most spheroid-dominated galaxies, 
present late bursts of SF associated to late (gaseous) mergers. This implies a formation 
scenario of spheroid-dominated galaxies in the field different to the expected for early-type galaxies 
in high-density environments.

$\bullet$ The cold gas velocity dispersion is moderately anisotropic. The $z=0$ vertical velocity dispersion profiles, $\sigma_z$($r$),
are nearly flat (except for the spheroid-dominated Sp5 galaxy), with values mostly around 25-35 \kms\ up to $\sim 1.5\rhalf$;
in several cases the dispersion increases with radius at larger radii. A model of self-regulated SF based
on the the vertical balance between the rate of kinetic energy injected by SNe/stellar winds and the rate of kinetic energy dissipation
by turbulence decay, is able to roughly predict the $\sigma_z$($r$) profiles. However, at radii where the
SF rate strongly decreased, the high  measured $\sigma_z$ values are not produced by SNe/stellar winds but
probably by gas accretion. The velocity dispersions 
of the simulated galaxies are significantly larger at higher redshifts, when the galaxies are actively assembling
and having strong bursts of SF. The average velocity dispersion histories of the simulated galaxies roughly 
follow the corresponding SF histories.

$\bullet$ The stellar mass growth efficiency, \ms/\mv, of our simulated galaxies at $z=0$
is in good agreement with semi-empirical inferences, showing that MW-sized systems
are in the peak of this efficiency. The evolution of \ms/\mv\ happens mostly
around the scatter of the present-day \ms/\mv--\mv\ correlation, which implies
that this correlation should not change significantly with $z$ in the $\mv=10^{11}-10^{12}$ \msun\
range, as some look-back time semi-empirical studies suggested.
For the disk-dominated runs, \ms/\mv\ and $M_b/\mv$ were very low
at the earliest epochs, then suddenly increased during the initial burst of SF, and when 
the halos overcome the mass of $\sim 5\times 10^{11}$ \msun, the galaxies entered in the
quiescent phase of stellar/baryonic mass growth nearly proportional to the cosmological halo mass
growth (\ms/\mv\ and $M_b/\mv \approx$ const. with time). 

$\bullet$ The galaxy baryonic fractions,  
$M_b/\mv$, are much lower than the universal one, $\Omega_b/\Omega_{m}$. 
Even taking into account the gas outside the galaxies, 
the missing baryons within \rv\ with respect to the universal fraction amounts for
 $\approx 25-50\%$. The baryonic budget within \rv\ for our eight MW-sized galaxies 
 shows that on average $\approx$ 28\% and 5.4\% of baryons are in stars and cold
 gas in the galaxy, and 3.4\%, 24.2\% and much less than 1\% are and in cool, warm-hot, and hot gas
 in the halo, respectively.
 

The main goal of our study was to show that a {\it simple} subgrid scheme, which
captures the main physics of SF and stellar feedback in a {\it effective} way, given our
resolution, is able to produce realistic galaxies formed in halos with masses during the evolution of
$\sim 10^{11}-10^{12}$ \msun. These galaxies end up today with total masses close
to that estimated for the MW galaxy. 
The success of this scheme can be attributed to the fact that 
at these scales neither the SF-driven 
outflows nor AGN-driven feedback, and the halo environmental effects (long gas 
cooling times), are so relevant as they are at the lower and higher
scales, respectively.  

\section*{Acknowledgements}   
The authors thank the anonymous Referee for insightful comments and helpful suggestions that
enriched the paper. VA  acknowledges CONACyT grant (Ciencia B\'asica) 167332 for partial support.

\end{document}